\documentclass[floatfix,aps,prx,groupedaddress,twocolumn]{revtex4-2}

\usepackage{CJK} 

\usepackage{xcolor}
\usepackage{epsfig}
\usepackage{relsize}
\usepackage{graphicx}
\usepackage{amsfonts}
\usepackage[figuresright]{rotating}
\usepackage{amssymb}
\usepackage{amsmath}
\usepackage{color}
\usepackage{tikz}
\usepackage{mathrsfs}

\DeclareGraphicsExtensions{.eps, .PNG, .jpg}

\newcommand{\ket}[1]{| #1 \rangle}
\newcommand{\bra}[1]{\langle #1 |}

\newcommand{\blcirc}[2][black,fill=black]{\tikz[baseline=-0.5ex]\draw[#1,radius=#2] (0,0) circle ;}%
\newcommand{\whcirc}[2][black,fill=white]{\tikz[baseline=-0.5ex]\draw[#1,radius=#2] (0,0) circle ;}

\begin{document}

\title{The fate of quantum many-body scars in the presence of disorder}
\author{Ian Mondragon-Shem,$^1$ Maxim G. Vavilov,$^2$ and Ivar Martin$^1$}
\affiliation{$^1$Material Science Division, Argonne National Laboratory, Argonne, IL 08540, USA}
\affiliation{$^2$Department of Physics, University of Wisconsin-Madison, Madison, Wisconsin 53706, USA}
\date{\today}
\begin{abstract}
Experiments performed on strongly interacting Rydberg atoms have revealed surprising persistent oscillations of local observables. These oscillations have been attributed to a special set of non-ergodic states, referred to as quantum many-body scars. Although a significant amount of research has been invested to understand these special states, it has remained unclear how stable scar states are against disorder.  We address this question by studying numerically and analytically the magnetization and spatio-temporal correlators of a model of interacting Rydberg atoms in the presence of disorder. While the oscillation amplitudes of these observables decay with time as the disorder strength is increased, their oscillation frequency remains remarkably constant.  We show that this stability stems from resonances in the disordered spectrum that are approximately centered at the same scar energies of the clean system. We also find that multiple additional sets of scar resonances become accessible due to the presence of disorder and further enhance the oscillation amplitudes. Our results show the robustness of non-ergodic dynamics in scar systems, and opens the door to understanding the behavior of experimentally realistic systems.
\end{abstract}

\maketitle

\section{Introduction}

Recent progress in the design and coherent control of programmable quantum simulators has made it possible to discover new and striking non-equilibrium phenomena \cite{Richerme2014,Jurcevic2014,Garttner2017,Bernien2017}. A particularly remarkable example has been the  detection of quantum many-body scars using a 51 qubit cold-atom platform \cite{Bernien2017}. In this system, strongly interacting Rydberg atoms were initially prepared in a product state and subsequently allowed to evolve under their own unitary dynamics. Unexpectedly, the dynamics of measured observables presented persistent oscillations at finite-energy density. This represented a violation of the Eigenstate Thermalization Hypothesis (ETH) \cite{Srednicki1994}, a central tenet of statistical mechanics. The ETH says that states at finite energy density evolve at long times into ergodic states wherein the expectation value of local observables converge to thermal values. The experimental observation of persistent oscillations thus signals that non-ergodic behavior is at play and that physics beyond the ETH was accessed by the Rydberg quantum simulator.

Subsequent analyses revealed that ETH was violated due to the presence of special states located throughout the energy spectrum which were termed many-body quantum scars \cite{Turner2018a,Turner2018b}. These states are embedded in an ergodic spectrum and are approximately equally spaced in energy, separated by the characteristic energy scale $\omega_{\text{scar}}.$ They exhibit less-than volume law entanglement and have high overlap with the initial product state used in the experimental simulation \cite{Bernien2017}. The observed persistent oscillations at the frequency $\omega_{\text{scar}}$ thus follow from these properties of quantum many-body scars. These states have garnered intense interest and have been found in a wide range of systems, both static and driven, as well as in higher dimensions \cite{Alhambra2020,ana2020quantum,Pai2019, Bull2019,Ok2019,mukherjee2020collapse,Mark2020,Zhao2020,Michailidis2020,Voorden2020,Lee2020}. Several insights have been  obtained by constructing exact scar eigenstates and from analytical results \cite{lin2019exact,Chattopadhyay2020,Mizuta2020,Moudgalya2020,moudgalya2020large,Iadecola2020}. Strategies have also been devised to understand and enhance the quality of persistent dynamics by recognizing emergent algebras in scar subspaces \cite{Choi2019,Bull2020,o2020tunnels}. Furthermore, fundamental questions about the nature of such scar states have been addressed by studying their proximity to integrability, as well as connections with concepts of quantum chaos \cite{Ho2019,Khemani2019,Christopher2020,Michailidis2020b}.

An important question that {\em has not}  yet been addressed is the effect of disorder on the stability of many-body scars. While disorder was discussed in \cite{Schecter2019, Shibata2020}, it was  only in the context of  fine-tuned models where disorder was designed not to affect the scar states themselves. The effects of disorder in the PXP model were studied in \cite{Chen2018}, but the focus was solely on many-body localization physics and not on the non-ergodic dynamics arising from quantum scars. Thus, the broader question of the effects of generic disorder on quantum scars has remained open.  One might expect that the randomizing effect of a disordered environment will generically suppress the oscillation amplitude and distort the oscillation frequency of observables. On the other hand, one could also expect that at least some quantum scars will be robust against disorder given  that strong kinematic constraints underlie their existence in some models.  These constraints introduce strong spatio-temporal correlations, suggesting enhanced robustness with respect to spatial imperfections. 

Undertaking this study is important for both practical and fundamental reasons.  As a practical matter, near-term quantum simulators, such as those based on superconducting qubits, are naturally affected by qubit-to-qubit variations \cite{Burnett2019,tannu2019not,Lisenfeld2019}. If quantum scars are to be probed in such platforms, it is necessary to understand how disorder will affect their detection. At the same time, disorder has played a central role in questions of ergodicity and non-equilibrium dynamics in many-body quantum systems.  The fundamental phenomenon of many-body localization (MBL) was predicted for sufficiently strong disorder in a general class of equilibrium interacting electronic systems \cite{Basko2006}.  Further studies of non-equilibrium phenomena in the presence of disorder have shown that many-body localization and associated phases of matter persist  in a wide variety of interacting system even far from equilibrium\cite{Oganesyan2007,Pal2010,Serbyn2013,Schreiber842,Nandkishore2015,Choi2016,Keyserlingk2016,Yao2017,Choi2017,Zhang2017}. In the space of models, systems with scars reside somewhere between the integrable and fully ergodic systems. It is therefore natural to expect  their response to disorder to yield a spectrum of unusual features, both in and out of equilibrium. This is indeed what we find in this work.

\begin{figure}
\centering
 \includegraphics[trim = 0mm 0.cm 0cm 0mm, clip,scale=0.38
]{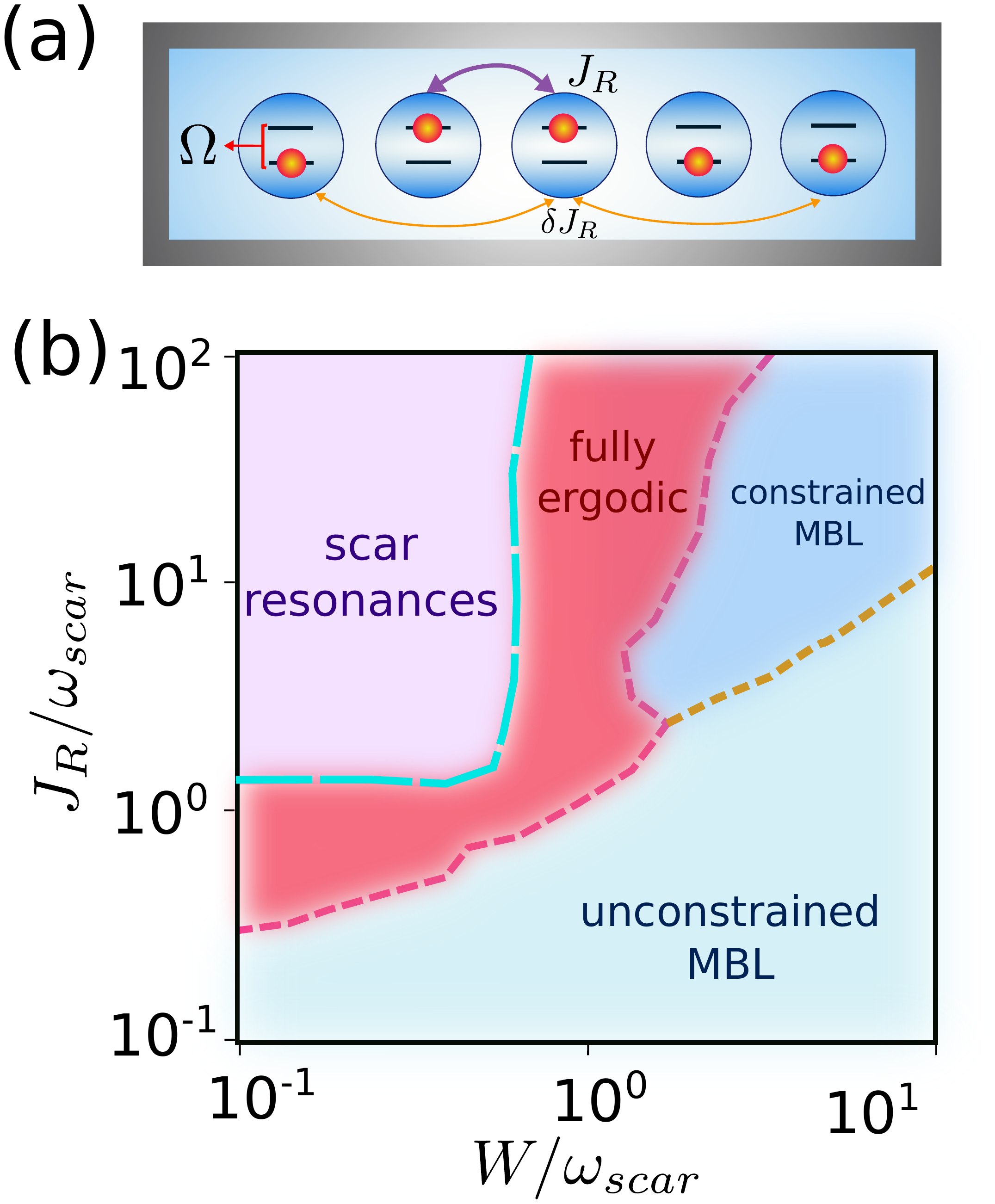}
\caption{\textbf{(a)} Schematic representation of the Rydberg simulator modeled by Eq.~(\ref{Eq_main}). The excited and ground states of each qubit are separated by an energy $\Omega.$ Only when two neighboring qubits are in their excited states will the qubits interact, with energy $J_R.$ When  $J_R\gg\Omega,$ a kinematic constraint develops that leads to the formation of quantum many-body scars. \textbf{(b)} Diagram of the dynamical regimes obtained in this work for a chain of  Rydberg atoms as a function of the nearest-neighbor interaction $J_R$ and disorder strengths   $W$: a regime that exhibits scar dynamics due to scar resonances, a regime that is ergodic without scar resonances, a constrained MBL phase, and an MBL without kinematic constraints.}\label{Fig_system}
\end{figure}

In this work we study how scars respond to generic disorder. We use the Rydberg Hamiltonian which accurately models the original quantum simulator where quantum scars were discovered \cite{Bernien2017} (the model system is schematically represented in Fig.\ref{Fig_system}a). We find that eigenstate diagnostics of scar states \cite{Turner2018a}  fail  when disorder is turned on.  However, clear oscillations can still be identified in observables such as the magnetization of the system as well as spatio-temporal correlators. While disorder indeed leads to the decay of these oscillations with time, their frequency  remains close to $\omega_{\text{scar}}$ for a finite range of disorder strengths, exhibiting significant rigidity to generic perturbations.

We provide a physical  picture for these observations by deriving analytically the disorder-averaged dynamics of the magnetization and temporal correlators in the weak-disorder limit. This derivation shows that, instead of remaining discrete eigenstates, scar states become transformed into {\em resonances} in the energy spectrum of the disordered system. For sufficiently small disorder strength, these resonances have widths smaller than $\omega_{\text{scar}}$, and are centered at the energies of the clean scars states. As a result, the observables we probe continue to oscillate at the scar frequency before decaying irreversibly. Surprisingly, we also uncover multiple additional scar resonances not associated with the previously known scar states; these novel states come into play in the presence of disorder, and, remarkably, enhance the magnetization oscillations at intermediate times. While in the bulk of this work we focus on the regime of strongly interacting Rydberg atoms (the so called ``PXP'' regime), at the end we fully map out the dynamical regimes of the Rydberg model for arbitrary coupling strength as a function of disorder.

Our results show that quantum many-body scars can be detected in moderately disordered quantum simulators. This in itself can potentially prove a valuable tool, for example, in the calibration of quantum hardware. Furthermore, we find that disorder is able to reveal the multi-tower structure of scar resonances in the Rydberg simulator. This insight serves as a new window into the underlying complexities of systems with scar states that could help further understand the circumstances under which they arise and how they can lead to ETH violation.

\subsection*{Outline}

We now outline the structure of this work and summarize the main results. We begin in Section \ref{Sec_Model} by introducing the model we use as a case study, namely the Rydberg Hamiltonian which describes interacting Rydberg-atom qubits; the system is shown schematically in Fig.\ref{Fig_system}a.  We discuss basic aspects of this model, including how scars arise in the strongly interacting limit, and also the type of disorder we consider in this work. 

In Section \ref{Sec_Disorder}, we consider the effects of disorder on the eigenstate properties of the PXP model. We discuss how scar states are no longer detectable in the energy spectrum (Fig.\ref{Fig_FailedDiagnostics}) and determine how the ergodicity of the system is affected (Fig.\ref{Fig_ergodicity}). We then move on in Section \ref{Sec_Magnetization} to carefully analyze the magnetization of the system. We show that it continues to carry signatures of non-ergodic oscillations at the scar frequency for moderate disorder strengths due to the kinematic constraint of the PXP model (Figs.\ref{Fig_ScarMagnetization},\ref{Fig_PXP_ParamVsPXP},\ref{Fig_PXP_flow}). 

In order to understand these results, in Section \ref{Sec_resonances} we derive explicitly the disorder-averaged magnetization dynamics in the weak disorder limit.  This derivation shows that the dynamics is controlled by disorder-induced resonances that arise from multiple towers of scar states of the clean system. Although their widths grow with disorder strength and this leads to the decay of the oscillation amplitude, non-ergodic oscillations continue to manifest in the magnetization as long as the width remains smaller than the scar frequency (Fig.\ref{Fig_MagnetizationComparison}).  We further show in Section \ref{Sec_Correlations} that scar resonances also account for temporal correlations in the system (Figs.\ref{Fig_ScarCorrelator},\ref{Fig_Z2correlator}).

We bring everything together in Section \ref{Sec_PD} by showing a diagram in the parameter space of interaction and disorder strengths which shows four distinct dynamical regimes (Fig.\ref{Fig_system}a): a regime with scar resonances, a regime that is ergodic and does not exhibit revival dynamics, a constrained many-body localized phase, and another many-body localized phase without kinematic constraints. Finally, in Section \ref{Sec_Conclusions} we discuss the main conclusions from this work, we comment on how our results can be tested in various quantum simulators, and propose future directions of research.

\section{Model}\label{Sec_Model}

In this section, we introduce the model that will serve as a case study to understand how quantum many-body scars respond to disorder.  We begin with the standard Rydberg Hamiltonian given by  \cite{Turner2018a,Turner2018b,Iadecola2019,Bull2019}
\begin{equation}
    H_{R}=\sum^L_{r=1}\left[\frac{\Omega}{2}\sigma^x_r+J_R P^+_r P^+_{r+1}\right],\label{Eq_main}
\end{equation}
where the Pauli matrices $\sigma^{x,y,z}_r$ act on ground and excited qubit states $\{\ket{\blcirc{2pt}},\ket{\whcirc{2pt}}\}$ respectively, we defined the projection operator $P^{\pm}_r=\frac{1}{2}\left(\mathbb{I}\pm \sigma^z_r\right),$ and we assumed periodic boundary conditions.  This Hamiltonian models a chain of Rydberg atoms, each driven resonantly with strength $\Omega$ (Rabi frequency) and interacting with strength $J_R$ when two nearest neighbors are in the excited state \cite{Saffman2010,Low2012}.

In the strongly interacting limit $J_R\gg \Omega,$ there is a large energy cost $J_R$ to go from the state $\ket{\cdots\,\whcirc{2pt}\,\blcirc{2pt}\,\whcirc{2pt}\,\whcirc{2pt}\cdots}$ to the state $\ket{\cdots\whcirc{2pt}\,\blcirc{2pt}\,\blcirc{2pt}\,\whcirc{2pt}\cdots},$ which has two neighboring qubits in their excited state. As a result, when $J_R\gg\Omega,$ the energy spectrum splits into $L$ bands. The Hilbert space spanned by each of these bands is characterized by a fixed number $n_{\text{exc}}$ of pairs of contiguous qubits in their excited state. When the system is initialized within one of these bands, the number $n_{\text{exc}}$ is thus preserved by the dynamics. 

In particular, the  product state $\ket{\mathbb{Z}_2}=\ket{\whcirc{2pt}\,\blcirc{2pt}\ldots\whcirc{2pt}\,\blcirc{2pt}},$ which we will use as the initial condition in this work, satisfies $n_{exc}=0.$  The effective Hamiltonian that generates the dynamics within this subspace, referred to as the PXP model \cite{Turner2018a}, is given by
\begin{equation}
    H_{pxp}= \frac{\Omega}{2}\sum^L_{r=1}\mathcal{P}\sigma^x_r\mathcal{P},\label{Eq_PXP}
\end{equation}
where $\mathcal{P}=\prod_{r}\left(\mathbb{I}-P^{+}_{r}P^{+}_{r+1}\right)$ projects into the $n_{exc}=0$ sector. The resulting Hamiltonian differs from that of a paramagnet by the presence of projection operators. To understand the impact on the dynamics of these projectors, we can express the Hamiltonian in the equivalent form $H_{pxp}= \frac{\Omega}{2}\sum^L_{r=1}P^{-}_{r-1}\sigma^x_rP^{-}_{r+1},$ which holds if we consider the dynamics of initial states that satisfy $\mathcal{P}\ket{\Psi(0)}=\ket{\Psi(0)}.$ This form of the Hamiltonian  illustrates that, while in a paramagnetic system the operator $\sigma^x_r$ generates rotations of the qubit at the position $r$ without regard to the state of other qubits, in the strongly interacting case this rotation can only happen when the qubits located at the positions $(r-1)$ and $(r+1)$ are in their respective ground states.

One striking consequence of this kinematic constraint is the large  (although incomplete) and periodic revival dynamics of product states such as the $\ket{\mathbb{Z}_2}$ state. Since this is a non-integrable model due to the kinematic constraint, the Eigenstate Thermalization Hypothesis (ETH) dictates that an initial state at finite-energy density should evolve under Eq.~(\ref{Eq_main}) into a thermal state at long times. However, the persistent and periodic revival of the $\mathbb{Z}_2$ state constitutes a direct violation of the ETH.  Spectrally, this can be understood in terms of a set of $L+1$ special eigenstates of the PXP model, referred to as quantum scars, which are embedded in a presumed ergodic spectrum within the $n_{\text{exc}}=0$ subspace. They are approximately equally spaced by an energy
\begin{equation}
 \omega_{\text{scar}}=2\pi\nu_{scar}= \eta \,\Omega, \quad \eta\approx0.636 . \label{Eq_omega}
\end{equation}
Scar stats have an exceptionally large overlap with the $\mathbb{Z}_2$ state, and exhibit less-than volume law spatial entanglement even though they reside at finite energy density. The properties of these special eigenstates explain the non-ergodic revival dynamics of the $\ket{\mathbb{Z}_2}$ state.

Now, the return probability of the $\mathbb{Z}_2$ state obtained by evolving with the PXP model is only partial. Furthermore, its spatial entanglement entropy grows with time, which suggests that the quantum scars of the conventional PXP model exhibit some level of hybridization with ergodic states. To correct for this, a suitable deformation $\delta H$ can be added to the Hamiltonian that removes this growth in entanglement entropy and further optimizes the return probability of an initial $\mathbb{Z}_2$ to reach unity \cite{Choi2019}. One can think of this ideal limit of optimized scars as a reference point for understanding the impact of disorder on its non-ergodic dynamics. We will use the particular deformation
\begin{equation}
    \delta H=\sum^L_{r=1}\delta J_R\left(\sigma^z_{r+2}+\sigma^z_{r-2}\right)\mathcal{P}\sigma^x_r\mathcal{P},
\end{equation}
where $\delta J_R=-0.026\Omega$ \cite{Choi2019}. Longer-ranged terms further improve the quality of scar states, but for our present purposes this deformation is sufficient.

The main subject of our work is the effect of disorder on the scar states, the corresponding revivals, and order parameter oscillations.
The disorder that we introduce is generic in that it breaks all the spin and translational symmetries 
 of the clean PXP model; namely, we consider the operator
\begin{equation}
    \hat{W}=\sum_{r,a}h_{a}(r)\sigma^a_r.\label{Eq_Dis}
\end{equation}
The random fields $h_{a}(r)$ with $a=x,y,z$ are chosen uniformly distributed in the range $[-W/2,W/2],$ where $W$ measures the strength of the disorder. The presence of all Pauli matrices breaks the symmetries that the clean PXP model has, such as chiral and time-reversal symmetries. 

Throughout most of this work we will mainly study the strongly interacting limit. This effectively projects the disorder operator into the $n_{exc}=0$ Hilbert space. As a result, the full Hamiltonian is given by
\begin{equation}
 H=H_{pxp}+\delta H+\mathcal{P}\hat{W}\mathcal{P}. \label{Eq_FullModel}
\end{equation}
Later, in Section \ref{Sec_PD}, we will lift this strong constraint to study the behavior of the system as a function of interaction strength $J_R$ in the full Hilbert space. 

When disorder is introduced into the system, the fields in Eq.~(\ref{Eq_Dis}) will tend to make the qubits in the system precess around random directions and at different rates. As a result, there will be a competition between the de-synchronizing effect introduced by the disorder potential and  the clean correlated oscillatory dynamics of the PXP Hamiltonian. Understanding the outcome of this competition is the main goal of this work.

\section{Spectral properties} 
\label{Sec_Disorder}

In this section, we study the impact of disorder on spectral and eingenstate properties of the Hamiltonian~(\ref{Eq_FullModel}) that are commonly attributed to the existence of non-ergodic dynamics of scar systems. We present evidence that scars can no longer be clearly identified as eigenstates of the system for moderate disorder strengths. Furthermore, we discuss how disorder initially enhances the ergodicity of the system at low disorder, and eventually induces a many-body localization transition at strong disorder. 

\subsection{Destruction of scar eigenstates}

\begin{figure}
\centering
 \includegraphics[trim = 0mm 0.cm 0cm 0mm, clip,scale=0.09]{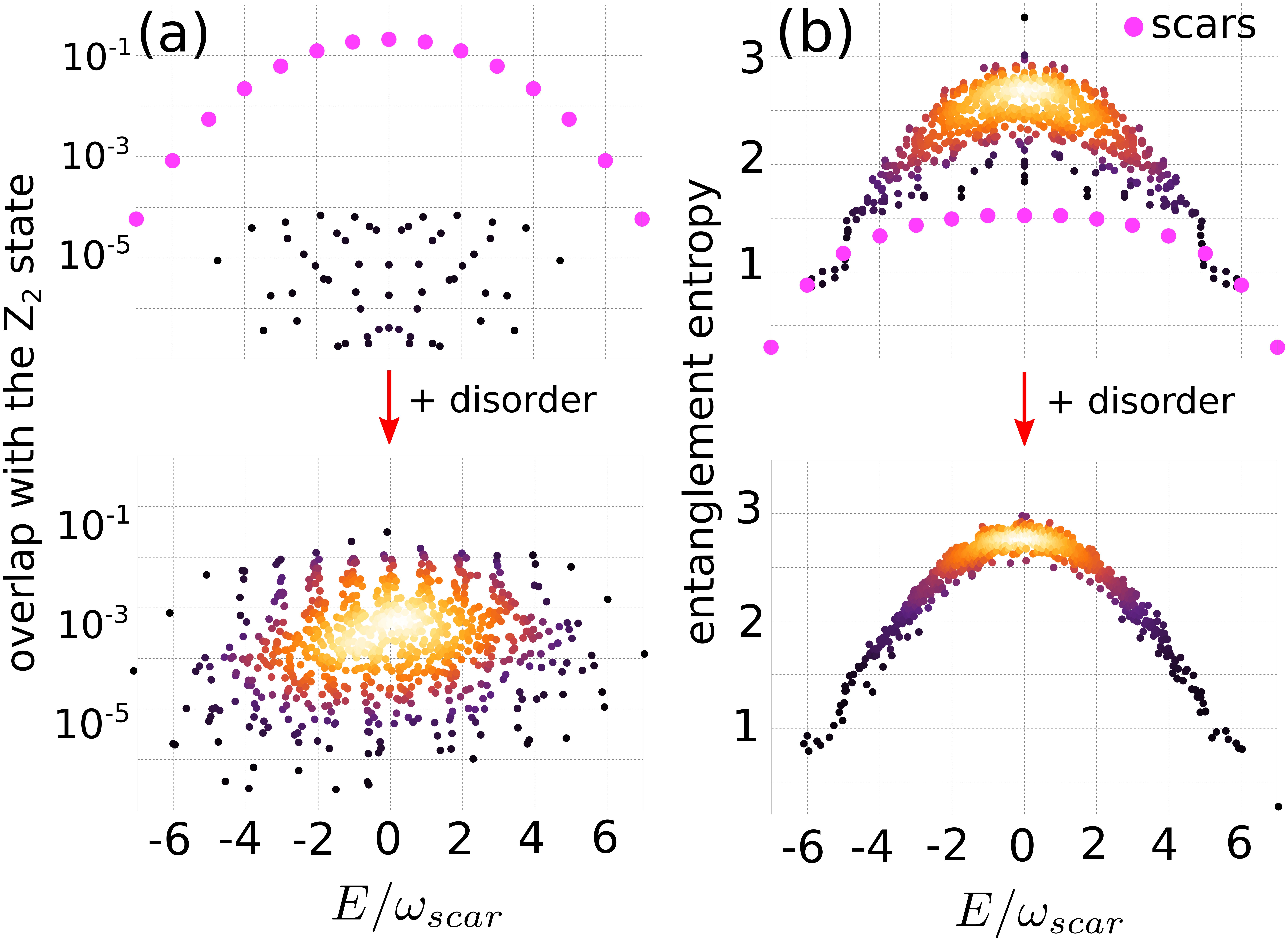}
\caption{ \textbf{(a)} Overlap with the $\mathbb{Z}_2$ state, and \textbf{(b)} entanglement entropy for each eigenstate with $L=14.$ Top figures correspond to the clean system, where clear signatures of scar states appear at the  integer values of $E/\omega_{\text{scar}}.$ The bottom figures correspond to disorder strength $W=0.25\omega_{\text{scar}}$ for a particular disorder realization. The scar signatures have become difficult to identify after disorder is introduced in the system.}\label{Fig_FailedDiagnostics}
\end{figure}

A basic question that first arises when disorder is introduced is whether scars can still be identified as eigenstates in the spectrum of the system. One standard way to identify scar states is to single out those eigenstates that have a high overlap the $\mathbb{Z}_2$ state \cite{Turner2018a,Turner2018b}. In Fig.\ref{Fig_FailedDiagnostics}a, we show the distribution of this overlap for $W=0$ and $W=0.25\omega_{\text{scar}}$ using a fixed disorder realization. While in the clean limit it is easy to identify states that have large overlap with the $\mathbb{Z}_2$ state, it is difficult to do so when disorder is turned on. Disorder has hybridized the clean-limit scar states with other spectrally near-by states to such an extent that the $\mathbb{Z}_2$ state no longer appears to be concentrated in a small subset of states. As a result, there are no clear signs of scar eigenstates in this overlap diagnostic.

Further evidence that scar eigenstates are absent in the spectrum can be obtained by calculating the spatial entanglement entropy of each energy eigenstate. If we divide the system in two halves $A$ and $B,$ the entanglement entropy of a state $\ket{E_n}$ is  $\mathcal{S}(E_n)=-\text{Tr}\left\{\rho^{(n)}_A\ln \rho^{(n)}_A\right\},$ where $\rho^{(n)}_A=\text{Tr}_{B}\left\{\ket{E_n}\bra{E_n}\right\}$ and $\text{Tr}_B$ traces the degrees of freedom in region $B.$ When the entanglement entropy is shown for each eigenstate in the system in the clean limit, as seen in Fig.\ref{Fig_FailedDiagnostics}b, it is possible to clearly discern isolated scar eigenstates that appear to have less-than volume law entanglement. Upon introducing disorder, however, the states of low entanglement recede quickly into the full set of volume-law ergodic states. Thus, we again see that hybridization of optimized scars with ergodic states has erased signatures of scar states in the entanglement of energy eigenstates.

\subsection{Enhanced ergodicity and localization transition}

In addition to obscuring signatures of scar eigenstates as we have seen in the previous section, disorder can also lead to other kinds of non-ergodic behavior such as many-body localization \cite{Pal2010,Nandkishore2015}. It is thus pertinent that we examine how the ergodicity of the system changes as the disorder strength is varied. To this end, we calculate the mean value of the level spacing ratios \cite{Oganesyan2007}
\begin{equation}
    r_n=\frac{\text{min}(\delta E_{n+1},\delta E_{n})}{\text{max}(\delta E_{n+1},\delta E_{n})},\label{Eq_rave}
\end{equation} 
where $\delta E_{n}=E_{n+1}-E_n$ with the energy eigenvalues $\{E_n\}$ sorted in ascending order. Since the disorder operator we have chosen does not respect any symmetries, the disordered Hamiltonian we are studying belongs to the Grand Unitary Ensemble (GUE), implying that $[\langle r_n \rangle]\approx 0.6$ when the system is ergodic \cite{Atas2013}, where $\langle \cdot \rangle$ denotes spectral averaging. When the system is localized, such as in an MBL phase, Poisson statistics dictates that $[\langle r_n\rangle]\approx 0.38.$

\begin{figure}
\centering
 \includegraphics[trim = 0mm 0.cm 0cm 0mm, clip,scale=0.36]{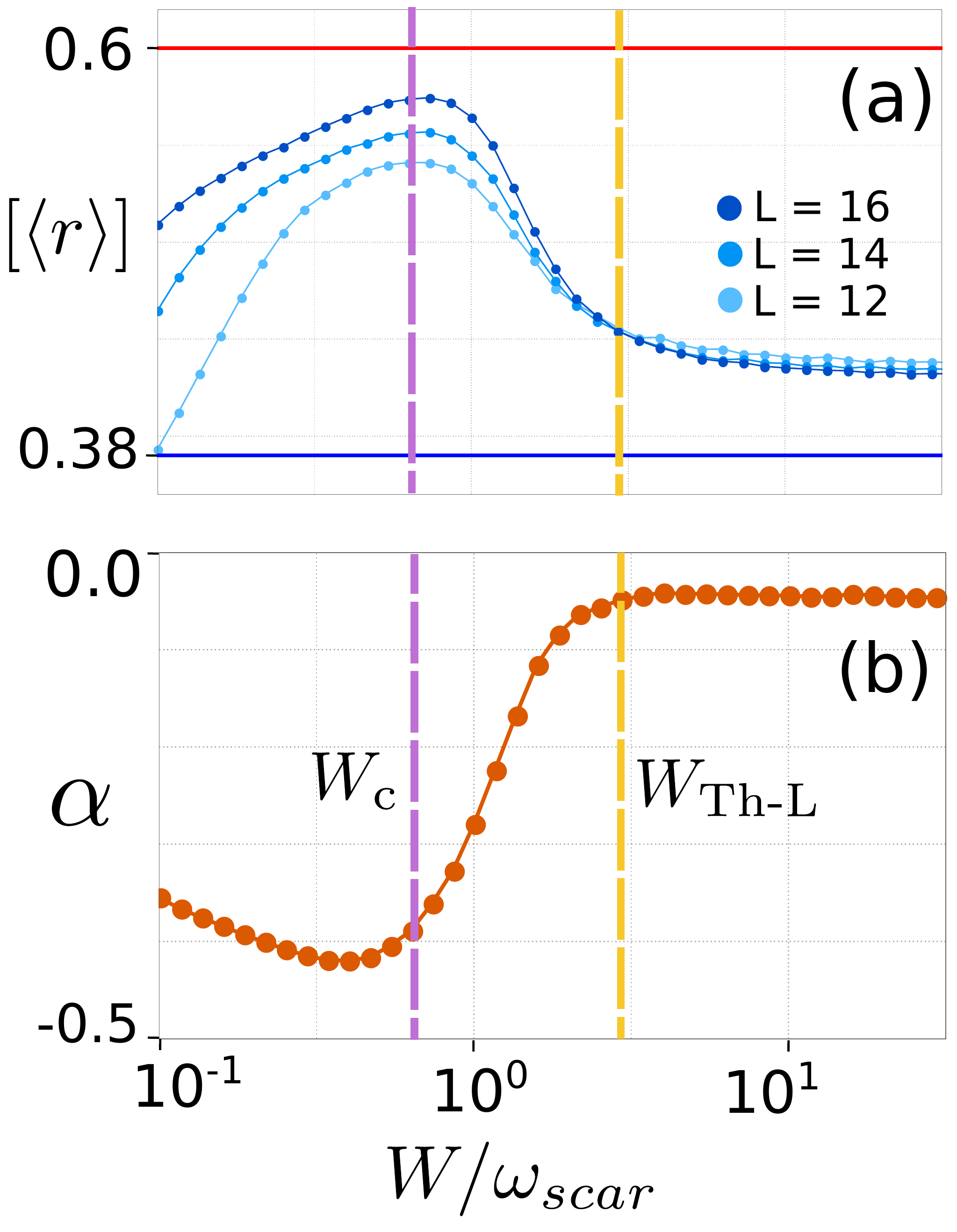}
\caption{\textbf{(a)} Mean level spacing ratios $[\langle r_n\rangle]$ defined by Eq.~(\ref{Eq_rave})  and \textbf{(b)} decay exponent of $\Delta \hat{\mathcal{O}}$ defined by Eq.~(\ref{Eq_Oop}) for the domain wall operator $\hat{\mathcal{O}} = \sigma^z_{L/2}\sigma^z_{L/2+1},$ as a function of disorder strength. The error bars in both figures are of the order or less than the circle markers. The disorder strength $W_c$ at which non-ergodic oscillations are lost is denoted by the purple line, whereas the disorder strength $W_{\text{Th-L}}$ at which the constrained MBL phase sets in is denoted by the yellow line.}\label{Fig_ergodicity}
\end{figure}

We present $[\langle r_n\rangle]$ in Fig.\ref{Fig_ergodicity}a as a function of system size. The flow with system size reveals signatures of a localization transition at around $W_{\text{Th-L}}\approx 2.13\omega_{\text{scar}}.$ When $W<W_{\text{Th-L}},$ the approach to the transition point does not occur monotonically for finite sizes.  Instead, the system presents an enhanced approach to the ergodic value $[\langle r_n \rangle]=0.6$ at a disorder strength somewhat below $W_{\text{Th-L}}.$   When $W>W_{\text{Th-L}},$ notice that $[\langle r_n\rangle]$ converges quite slowly to the Poisson value at strong disorder, which is the value that is expected for a many-body localized phase.  This is clearly due to the kinematic constraint imposed by the presence of projection operators.  Slow convergence to the Poisson value was also obtained in \cite{Chen2018} for a similar disordered PXP model, where this type of many-body localization under kinematic constraints was dubbed constrained MBL.

\begin{figure*}
\centering
 \includegraphics[trim = 0mm 0.cm 0cm 0mm, clip,scale=0.18]{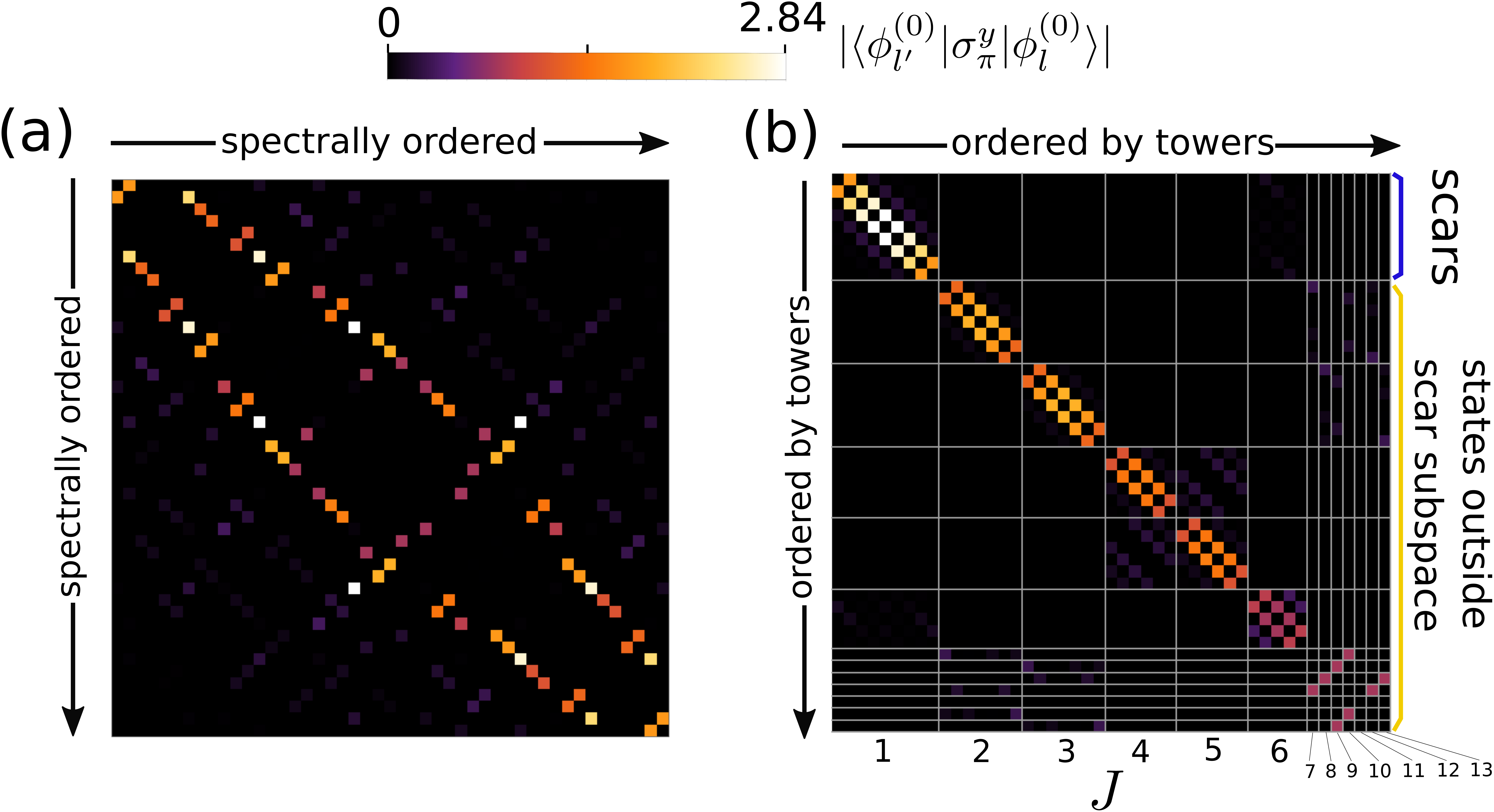}
\caption{ Density plot of the matrix elements $\vert\bra{\phi^{(0)}_{l'}}\widetilde{\sigma}^{y}_{\pi}\ket{\phi^{(0)}_{l}}\vert$ with respect the basis of energy eigenstates ordered by \textbf{(a)} energy  and \textbf{(b)} towers, respectively, for $L=8.$ The corresponding figure for  $\vert\bra{\phi^{(0)}_{l'}}\widetilde{\sigma}^{z}_{\pi}\ket{\phi^{(0)}_{l}}\vert$ is similar.}\label{Fig_towers}
\end{figure*}

Since level statistics at low disorder can be affected by the symmetries of the clean system, we also compute an additional diagnostic of ergodicity that makes use of the eigenstates of the system to confirm these results. The ETH states that the expectation value of local observables $\hat{\mathcal{O}}$ varies smoothly with energy. As a result, contiguous states satisfy
\begin{equation}
    \Delta \mathcal{O} = \vert\bra{E_n} \hat{\mathcal{O}}\ket{E_n}-\bra{E_{n-1}} \hat{\mathcal{O}}\ket{E_{n-1}}\vert \sim e^{-\mathcal{S}(E_n)/2},\label{Eq_Oop}
\end{equation}
where $\mathcal{S}(E)$ is the entropy at energy $E.$ A particular choice one can consider is the domain wall operator $\hat{\mathcal{O}} = \sigma^z_{L/2}\sigma^z_{L/2+1}.$ This operator was used to study signatures of integrability in the clean PXP model \cite{Khemani2019}. Near infinite temperature, we have $\mathcal{S}(E)\sim \ln\mathcal{D},$ where $\mathcal{D}$ is the dimension of the Hilbert space. This implies that $\Delta \hat{\mathcal{O}}\sim \mathcal{D}^{-1/2}$ if the system is ergodic. We proceed as in \cite{Khemani2019}, and we calculate $\Delta \hat{\mathcal{O}}\sim \mathcal{D}^{\alpha}$ as a function of $\mathcal{D}$ to extract the exponent $\alpha.$  For ergodic systems, the exponent $\alpha$ should thus converge to $-1/2,$ whereas in the non-ergodic regime we expect $\Delta \hat{\mathcal{O}}$ to decay slower with $\vert\alpha\vert<1/2.$ We show the exponent $\alpha$ extracted by using $L=12,14,16$ in Fig.\ref{Fig_ergodicity}b. The exponent approaches $-1/2$ below $W_{\text{Th-L}},$ and is significantly suppressed above it, consistent with our findings using energy level statistics.

\section{System dynamics}\label{Sec_Magnetization}

The results from the previous section raise the question of whether non-ergodic dynamics can still be detected in the presence of disorder. To explore this question, we turn to the study of the magnetization of the system. We define the on-site magnetization as
\begin{equation}
\mathcal{M}_a(r,t)=\bra{\Psi(t)}\sigma^a_r\ket{\Psi(t)},
\end{equation}
where $\ket{\Psi(t)}$ is the time-evolved state obtained from solving the Schrodinger equation with the initial condition $\ket{\Psi(0)}=\ket{\mathbb{Z}_2}.$ We further define the Fourier transforms  $\widetilde{\mathcal{M}}_a(k,t)=\bra{\Psi(t)}\widetilde{\sigma}^a_k\ket{\Psi(t)},$ $\overline{\mathcal{M}}_a(k,\omega)=\int^{\infty}_0 dte^{-i\omega t}
\widetilde{\mathcal{M}}_a(k,t)$, where
\begin{equation}
\widetilde{\sigma}^a_k=\sum^L_{r=1}e^{ikr}\sigma^a_r.    
\end{equation}
In what follows, we will first calculate the magnetization in the clean limit in order to understand how it depends on the presence of scar eigenstates. We will then investigate its behavior when disorder is introduced.

\begin{figure*}
\centering
 \includegraphics[trim = 0mm 0.cm 0cm 0mm, clip,scale=0.28]{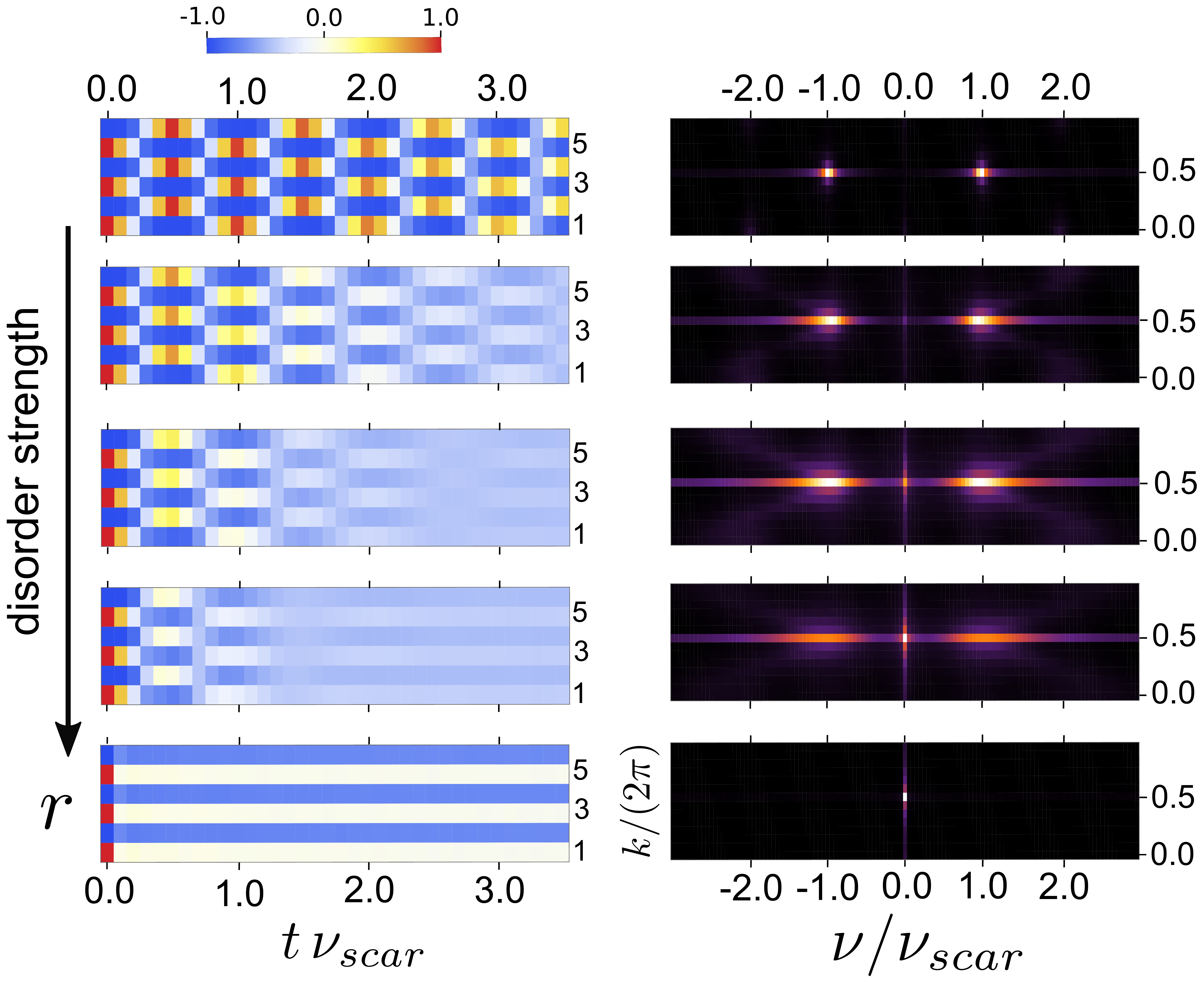}
\caption{\textbf{Left:} time evolution of $[\mathcal{M}_z(r,t)]$ for disorder strengths  $W/\omega_{\text{scar}}=0.1, 0.19, 0.63, 1.58, 7.41.$ In these figures, $L=16,$ but we only show the first six sites for visual clarity. \textbf{Right:} corresponding disorder-averaged Fourier component $[\vert\overline{\mathcal{M}}_z(k,\omega)\vert]$ at the same disorder strengths. For weak disorder, oscillations occur at $(k,\omega)=(\pi,\omega_{\text{scar}}).$ As the disorder strength is increased, the distribution broadens, and eventually concentrates at $(k,\omega)=(\pi,0).$}\label{Fig_ScarMagnetization}
\end{figure*}

\subsection{Scar signatures in the clean limit}

Let us begin by writting the initial state as $\ket{\Psi(0)}=\sum_{l}\psi_l \ket{\phi^{(0)}_l},$ where $\{\ket{\phi^{(0)}_l}\}$ is the set of eigenstates of the clean system. Since the time-evolved state is given by $\ket{\Psi(t)}=\sum_{l}e^{-iE^{(0)}_l t}\psi_l\ket{\phi^{(0)}_l},$ the magnetization takes the form
\begin{equation}
\widetilde{\mathcal{M}}_a(k,t)=\sum_{l,l'}e^{i\omega_{l'l}t}\psi^*_{l'}\psi_{l}\bra{\phi^{(0)}_{l'}}\widetilde{\sigma}^a_{k}\ket{\phi^{(0)}_l},\label{Eq_GenMag}
\end{equation}
where we defined $\omega_{l'l}=E^{(0)}_{l'}-E^{(0)}_{l}.$ The calculation of the magnetization hinges on understanding the behavior of the matrix elements $\bra{\phi^{(0)}_{l'}}\widetilde{\sigma}^a_{k}\ket{\phi^{(0)}_l}.$ 

Although the behavior of these matrix elements is not known for arbitrary wave vector $k,$ progress can be made for the particular case when $k=\pi$ and the initial state is $\ket{\Psi(0)}=\ket{\mathbb{Z}_2}$ \cite{Choi2019,Iadecola2019}. Given that the $\mathbb{Z}_2$ state is approximately spanned by optimized scar states, we only need to consider matrix elements $\bra{\phi^{(0)}_{l'\in \Lambda_s}}\sigma^{a}_{\pi}\ket{\phi^{(0)}_{l\in\Lambda_s}},$ where $\Lambda_s$ is the set of indices of optimized scar states. In particular, approximate ladder operators $\sigma^{\pm}_{\text{s}}$ can be defined within the scar subspace where
\begin{equation}
    \sigma^{\pm}_{\text{s}}=\frac{1}{2}\left(\widetilde{\sigma}^z_{\pi}\mp i\eta^{-1} \widetilde{\sigma}^y_{\pi}\right),
\end{equation}
which connect scar states that differ by an energy $\pm \omega_{\text{scar}}$ (see Appendix \ref{App_0} for details). The matrix elements $\bra{\phi^{(0)}_{l'\in \Lambda_s}}\sigma^{\pm}_{\text{s}}\ket{\phi^{(0)}_{l\in\Lambda_s}}$ that contribute to the magnetization are those for which $\omega_{l'l}=\pm \omega_{\text{scar}}.$ Plugging this into Eq.~(\ref{Eq_GenMag}) with $k=\pi$ leads to the magnetization components
\begin{subequations}
\begin{eqnarray}
    \widetilde{\mathcal{M}}_y(\pi,t)&\approx&\eta L\sin\left(\omega_{\text{scar}}t\right),\\
    \widetilde{\mathcal{M}}_z(\pi,t)&\approx&-L \cos\left(\omega_{\text{scar}}t\right),
\end{eqnarray}\label{Eq_MScarEx}
\end{subequations}
which exhibit oscillations at the scar frequency with fixed amplitude. These magnetization oscillations capture the non-ergodic dynamics induced by the presence of scars in the spectrum, and could thus be useful observables to track any remaining signatures of scar physics when the system becomes disordered.

Now, when disorder is introduced, other states of the clean system  outside of the scar subspace will inevitably become populated. As a result, in the following sections we will need to also understand the behavior of matrix elements $\bra{\phi^{(0)}_{l'\not\in \Lambda_s}}\sigma^{\pm}_{\text{s}}\ket{\phi^{(0)}_{l\not\in\Lambda_s}}.$ Not much has been said in the literature, however, about their behavior. These states have been largely assumed to be ergodic, which might suggest that they do not contribute to oscillations at the scar frequency.  Consider, however, Fig.\ref{Fig_towers}a, where we show a density plot of all the matrix elements $\vert\bra{\phi^{(0)}_{l'}}\widetilde{\sigma}^{y}_{\pi}\ket{\phi^{(0)}_{l}}\vert,$ with $l,l'$ ordered according to the energy of the corresponding state (we included the set of optimized scars). The structure of this matrix reveals that, while most entries are vanishingly small, there are some that are exceptionally large. At first glance there is no apparent order to the distribution of these large matrix elements. However, a clear pattern emerges when one performs an appropriate re-ordering of the energy basis by grouping them into sets that are connected via the same ladder operators $\sigma^{\pm}_{\text{s}}.$ This re-ordered basis reveals that the energy eigenstates can be organized into towers labeled by an index $J,$ with each tower containing a number $\mathcal{D}_J$ of states labeled by an index $m.$ There are a few states near the middle of the spectrum that do not seem to form towers with any other state, so they form sets with a single state each; such sets do not contribute appreciably to the magnetization. 

Using this reordering, we again show $\vert\bra{\phi^{(0)}_{J'm'}}\widetilde{\sigma}^{y}_{\pi}\ket{\phi^{(0)}_{Jm}}\vert$ in Fig.\ref{Fig_towers}b (details of how we performed this re-ordering are presented in Appendix \ref{App_Reorg}). The matrix is approximately block-diagonal with respect to $J,$ and each block is tri-diagonal, so we approximately write
\begin{equation}
    \bra{\phi^{(0)}_{J'm'}}\widetilde{\sigma}^{a}_{\pi}\ket{\phi^{(0)}_{Jm}}\approx\delta_{J'J}\left(\Gamma^{a,+1}_{Jm}\delta_{m' m\pm 1}+\Gamma^{a,-1}_{Jm}\delta_{m' m-1}\right),\nonumber
\end{equation} 
for $a=y,z.$ In this expression, we defined $\Gamma^{a,\kappa}_{Jm}=\bra{\phi^{(0)}_{Jm+\kappa}}\widetilde{\sigma}^{a}_{\pi}\ket{\phi^{(0)}_{Jm}}.$  Similar to the case of optimized scars (which in our notation corresponds to the $J=1$ tower), this tri-diagonal structure selects frequencies in   Eq.~(\ref{Eq_GenMag}) that are close to $\omega_{\text{scar}}$ for most towers.  As we will see in the following section, upon introducing disorder, these additional towers will become populated and their non-ergodic oscillations near the scar frequency will be revealed.   

We note that it is remarkable that the PXP model is comprised of multiple towers of scar states, of which the set of optimized scar states is just one example. One of the reasons that such towers have remained hidden is, in part, because the $\mathbb{Z}_2$ state is spanned largely by only the $J=1$ tower.  The multi-tower structure of the PXP model can lead to non-ergodic $SU(2)$ dynamics with a variety of product states as we discuss elsewhere \cite{MS2020}.

\begin{figure}
\centering
 \includegraphics[trim = 0mm 0.cm 0cm 0mm, clip,scale=0.46]{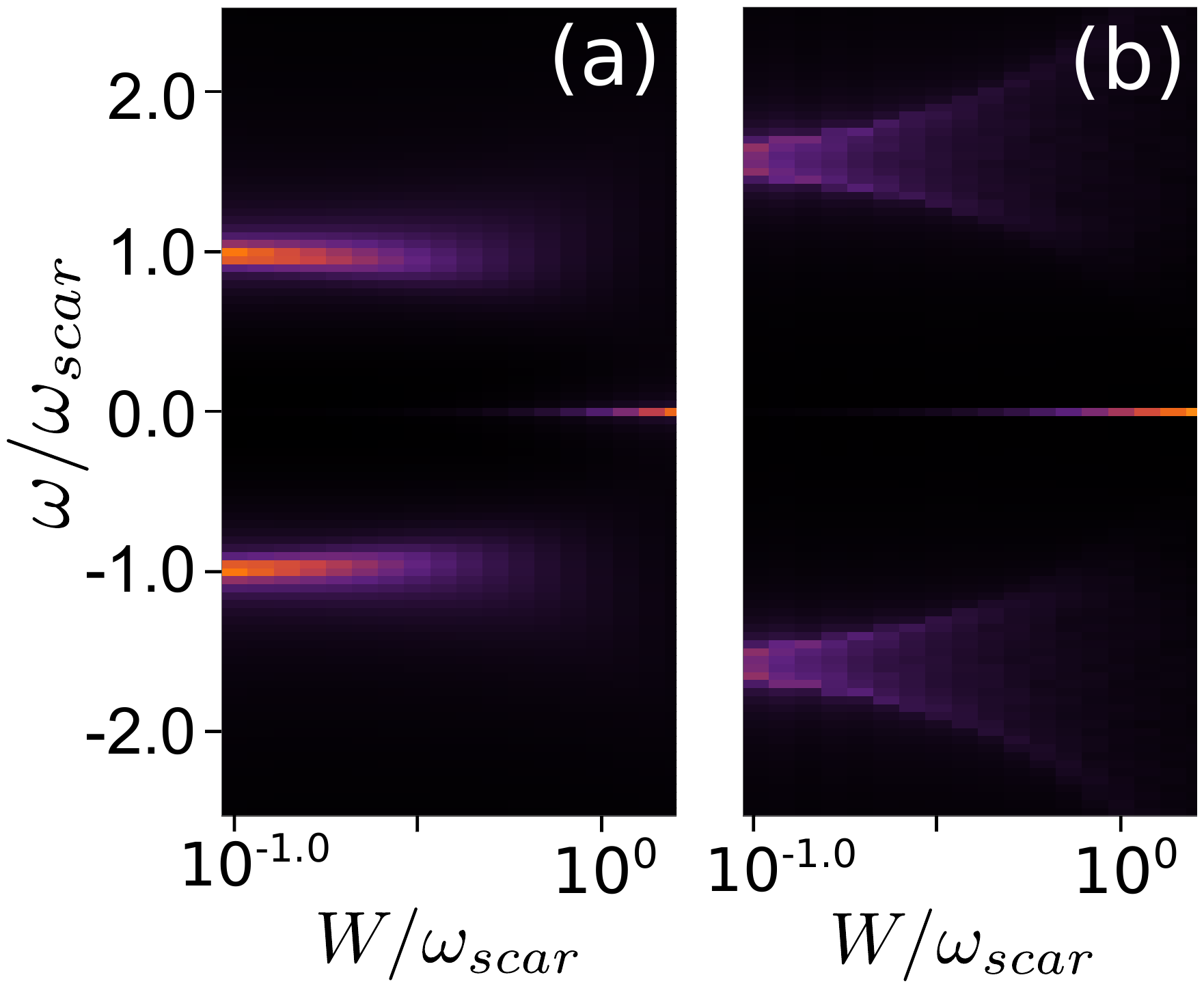}
\caption{Fourier component $[\vert \overline{\mathcal{M}}_z(\pi,\omega)\vert]$ as a function of $\omega$ and $W$ for the PXP model \textbf{(a)} and the free paramagnet \textbf{(b)}. A  pronounced broadening occurs in the paramagnetic case, whereas in the PXP model the dynamics is consistently centered and localized at the scar frequency.}\label{Fig_PXP_ParamVsPXP}
\end{figure}

\subsection{Stability of oscillations at the scar frequency}

Let us now examine the stability of magnetization oscillations at the scar frequency. We begin by calculating the on-site magnetization $[\mathcal{M}_z(r,t)]$ and its Fourier distribution $[\vert\overline{\mathcal{M}}_z(k,\omega)\vert]$ for a few  values of the disorder strength, as shown in Fig.\ref{Fig_ScarMagnetization}. Here, and throughout this work, all disorder-averaged quantities shown are averaged over $500$ disorder realizations and are denoted by square brackets $[\cdot].$ As can be seen in Fig.\ref{Fig_ScarMagnetization}, for the weak disorder case $W=0.1\omega_{\text{scar}},$ there is a clear periodic pattern in both space and time. Temporally, each qubit oscillates as a function of time at the scar frequency, initially between the maximum amplitudes $[\mathcal{M}_z(r,t)]=\pm1$ and gradually decaying with time. Spatially, the even and odd sites appear anti-ferromagnetically ordered throughout the dynamics, as expected from the kinematic constraint of the PXP model. This leads to sharp peaks in $[\vert\overline{\mathcal{M}}_z(k,\omega)\vert]$ at $(k,\omega)=(\pi,\pm \omega_{\text{scar}})$ as shown in the right column of  Fig.\ref{Fig_ScarMagnetization}, consistent with Eq.~(\ref{Eq_MScarEx}).  As the disorder strength is increased, the magnetization amplitude decays faster in time, until there is no longer any clear oscillation at the scar frequency for sufficiently strong disorder. Correspondingly, the Fourier peaks broaden, especially in the frequency domain. It is clear, however, from both $[\mathcal{M}_z(r,t)]$ and $[\vert\overline{\mathcal{M}}_z(k,\omega)\vert],$ that the frequency of oscillation continues to be dominated by $\omega=\omega_{\text{scar}}$ for a finite  range of disorder strengths even though the qubits are experiencing random and biased fields. This indicates that the dynamics of the system continues to exhibit signatures of many-body scars. For strong enough disorder, the distribution eventually focuses at $(k,\omega)=(\pi,0),$ indicating that the dynamics of the system has been rendered temporally trivial while maintaining antiferromagnetic ordering.

In order to better understand how stable the oscillations remain at the scar frequency, it is useful to compare $[\vert\overline{\mathcal{M}}_z(\pi,\omega)\vert]$ for both the deformed PXP and the paramagnetic case $J_R=0.$ The paramagnetic limit is equivalent to removing the projection operators from the PXP model. Because of this, the contrast between the two systems can shed light on the possible stabilizing effects that the kinematic constraint has on the oscillations of the scar system. In Fig.\ref{Fig_PXP_ParamVsPXP}a,b we show $[\vert\overline{\mathcal{M}}_z(\pi,\omega)\vert]$ for both limits as a function of disorder strength. A sharp difference is revealed in their response to the presence of disorder. In the paramagnetic case, the dominant frequency of oscillation at the lowest disorder strength $W=0.1\,\omega_{\text{scar}}$ we show in this figure is centered at $\omega=\Omega,$ although it is already visibly broadened at this disorder strength. As $W$ is increased, the broadening grows continuously, to the point that there is no clear dominant frequency of oscillation. As expected, the ease with which the dynamics develops a broad range of frequencies is the manifestation of each qubit oscillating at its own local disorder-induced Zeeman field.

\begin{figure}
\centering
 \includegraphics[trim = 0mm 0.cm 0cm 0mm, clip,scale=0.33]{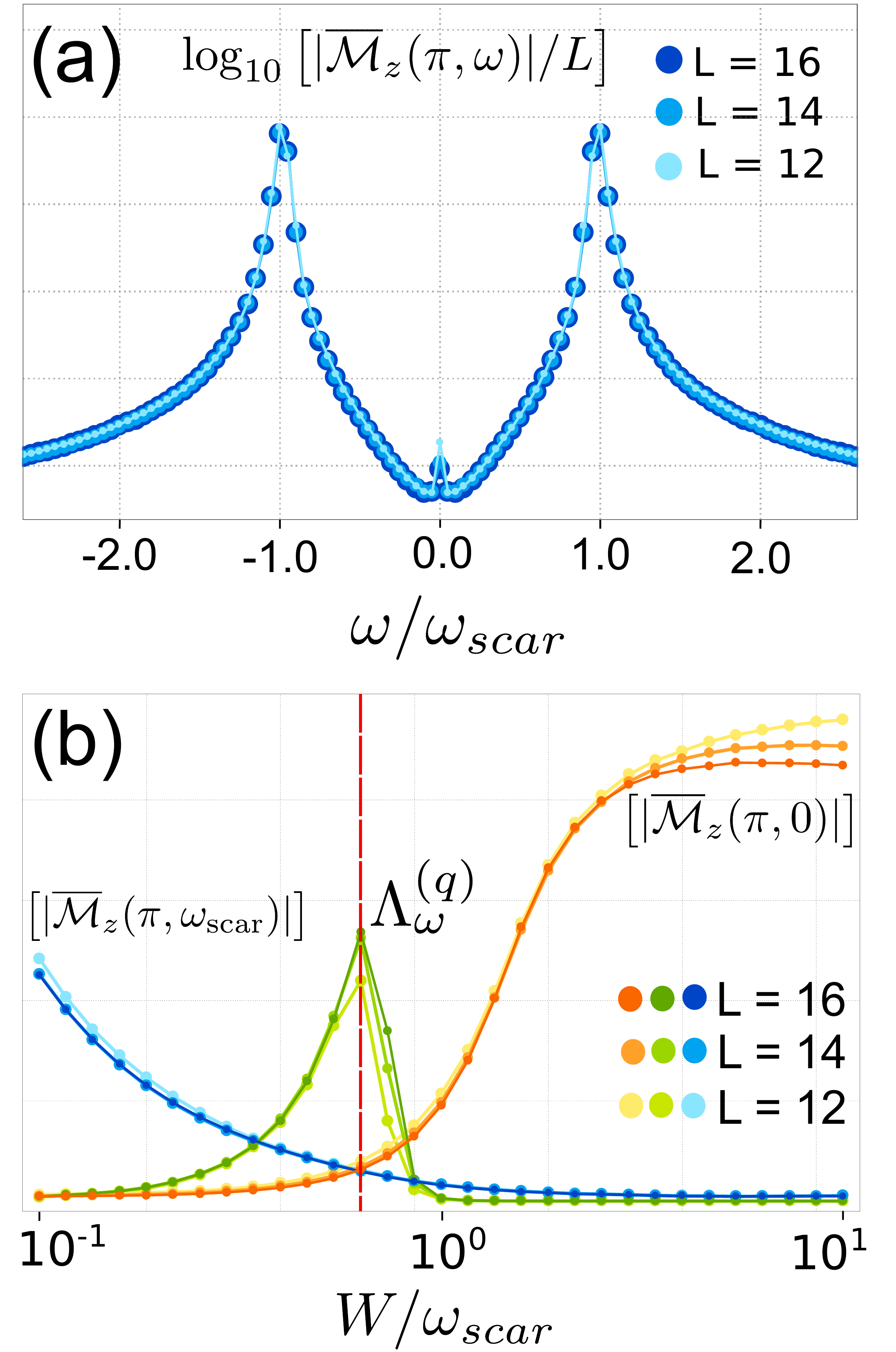}
\caption{\textbf{(a)} Behavior of $\log_{10}[\vert \overline{\mathcal{M}}_z(\pi,\omega)\vert/L]$ as the system size is increased with $W=0.1 \omega_{\text{scar}},$ showing that there are no appreciable changes.  \textbf{(b)} Behavior of $[\vert \overline{\mathcal{M}}_z(\pi,\omega_{\text{scar}})\vert],$ $[\vert \overline{\mathcal{M}}_z(\pi,0)\vert]$  and $\Lambda^{(q=6)}_{\omega}$ as a function of disorder strength, for three system sizes. There is a clear and stable transition between a regime with scar dynamics and a dynamically trivial regime. In this figure, we chose $q=6$ for convenience as it accentuates the position of the peak, but other values of $q$ also work. }\label{Fig_PXP_flow}
\end{figure}

By contrast, in the strongly interacting case there is a clear persistent dominant frequency of oscillation centered around $\omega=\omega_{\text{scar}}.$  This occurs in spite of the qubits experiencing the same random fields as those used in the paramagnetic case. The strong kinematic constraint that correlates the rotation of a given qubit with respect to its neighbors continues to robustly enforce oscillations at the scar frequency. As a check of the stability of this observation, we find that the oscillations at the scar frequency persist when the system size is increased, as we exemplify  in Fig.\ref{Fig_PXP_flow}a, where  $[\vert\overline{\mathcal{M}}_z(\pi,\omega)\vert]/L$ does not change appreciably for $L=12,14,16$ with $W=0.1\omega_{\text{scar}}.$ We thus find clear evidence that the interacting system continues to sustain oscillations at the scar frequency for a significant range of disorder strengths, even when conventional eigenstate diagnostics fail.

Oscillations at the scar frequency terminate at a characteristic disorder strength $W_c.$ This can be seen in Fig.\ref{Fig_PXP_ParamVsPXP}a, where there appears to be a marked change in dynamical regimes wherein a strong signal in $[\vert\widetilde{\mathcal{M}}_z(\pi,\omega)\vert]$ eventually yields to a clear peak at zero frequency when the disorder strength is increased. To better visualize this, in Fig.\ref{Fig_PXP_flow}b we show  $[\vert\widetilde{\mathcal{M}}_z(\pi,\omega_{\text{scar}})\vert]$ and $[\vert\widetilde{\mathcal{M}}_z(\pi,0)\vert]$ as a function of disorder strength. At weak but non-zero disorder, the dynamics is dominated by $[\vert\widetilde{\mathcal{M}}_z(\pi,\omega_{\text{scar}})\vert].$ As the disorder strength is increased, the peak at $\omega_{\text{scar}}$ gradually decreases, while at the same time the zero frequency component increases, eventually rendering the time evolution trivial at strong disorder. The change between both regimes can be taken to occur when both frequency components have the same magnitude, namely round $W_{c}\approx 0.63 \omega_{\text{scar}}$. As can be seen in this figure, the crossing between both curves does not appear to change as the system size is increased. 

To capture this transition more clearly, we make use of the normalized frequency distribution \cite{Choi2017,Rovny2018} $\mathcal{F}(\omega)=\frac{1}{\mathcal{N}}[\vert\widetilde{\mathcal{M}}_z(\pi,\omega)\vert],$ where $\mathcal{N}=\sqrt{\int d \omega[\vert\widetilde{\mathcal{M}}_z(\pi,\omega)\vert]^2}.$ 
We introduce the quantity $\Lambda^{(q)}_{\omega}=\left[\int d \omega\mathcal{F}^q(\omega)\right]^{-1},$ which measures the spread of $\mathcal{F}(\omega)$ in the frequency domain. This is analogous to the participation ratio defined in real-space to study the spatial localization of  wave functions.

As we show in Fig.\ref{Fig_PXP_flow}b,  $\Lambda^{(q=6)}_{\omega}$ is suppressed at both weak and strong disorder, as expected since the system is strongly dominated by $\omega=\omega_{\text{scar}}$ and $\omega=0,$ respectively. At intermediate disorder strengths, $\Lambda^{(q)}_{\omega}$ develops a peak at $W_c$ that does not appear to shift significantly with system size. This peak is present for other values of $q\neq 6$ as well, so our particular choice for $q$ is somewhat arbitrary. It is used in this work because it yields a sharper peak than those obtained with lower values of $q$ (higher values can make it even sharper).  The peak signals a spread of the frequency components at $W_c,$ allowing us to identify two distinct dynamical regimes, one with clear oscillations at the scar frequency and another that is manifestly featureless.

These results thus show that oscillations at the scar frequency persist over an appreciable time scale even as disorder is making the system more ergodic. It is clear that the increased ergodicity induced by disorder is consistent with the temporal decay of the magnetization amplitude. What is nontrivial is that these oscillations remain close to $\omega_{\text{scar}},$ indicating temporal rigidity. 

Given that the characteristic disorder strengths $W_{c}$ and $W_{\text{Th-L}}$ appear to be distinct (as indicated in Fig.\ref{Fig_ergodicity}), there are three distinct dynamical regimes that arise as a function of disorder strength: a regime with oscillations at the scar frequency, a fully thermal phase, and a constrained MBL phase. In the following section, we will present a quantitative interpretation for these observations that will show that oscillations at the scar frequency occur due to the presence of scar resonances in the spectrum.

\section{Scar resonances}\label{Sec_resonances}

The behavior of the magnetization poses a number of questions regarding the presence of scars in the disordered system. Scars no longer exist as eigenstates in the spectrum, and yet the magnetization shows multiple oscillations in an increasingly ergodic system before decaying at long times. Central to the underlying dynamics is the fact that disorder couples optimized scars with the rest of the spectrum. This brings into focus, on the one hand, the manner in which scars hybridize with nearby ergodic states and, on the other, the role that states outside of the scar subspace play in the decay of the magnetization signal.

\begin{figure*}
\centering
 \includegraphics[scale=0.4]{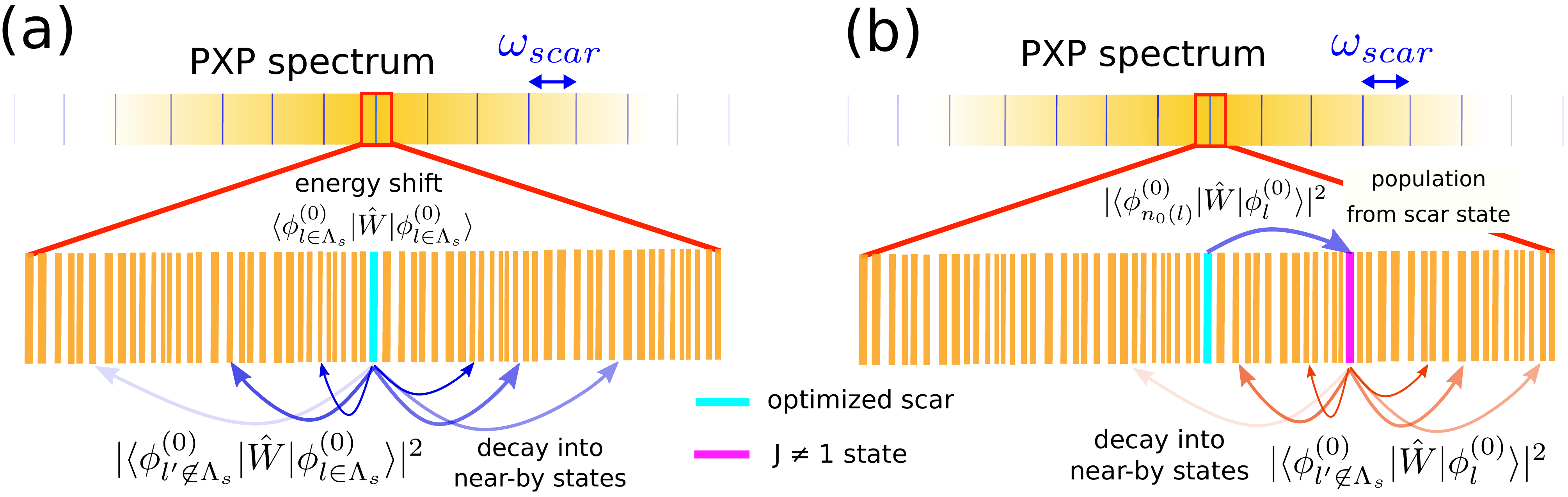}
\caption{\textbf{(a)} Direct decay of scar states.  The disorder induces transitions from the scar state into spectrally nearby generic states,  with the rate  controlled by matrix elements of the form $\vert\bra{\phi^{(0)}_{l'\not\in\Lambda_s}} \hat{W}\ket{\phi^{(0)}_{l\in\Lambda_s}}\vert^2$. Additionally, there is an energy shift of the form $\bra{\phi^{(0)}_{l\in\Lambda_s}} \hat{W}\ket{\phi^{(0)}_{l\in\Lambda_s}}.$ \textbf{(b)} Decay via $J\ne 1$ tower states. The initial scar state couples  (blue arrow) to a  state in a separate tower (purple line), which then decays (orange arrows) into the spectrally nearby generic states. These processes give significant contribution (comparable to the main channel in \textbf{(a)}) to the order parameter dynamics at finite times. }\label{Fig_DecayPic}
\end{figure*}

As we will see in this section, the robustness in the oscillations occurs in fact via two mechanisms. First, we will present evidence that scars continue to exist in the system in the form of resonances with a nonzero width. As time goes by, the finite life time of these resonances leads to their decay into other states in the spectrum. Additionally, although the states outside the scar subspace are more highly entangled and ergodic, they also lead to oscillations close to the \textit{same} scar frequency, contributing appreciably to the magnetization of the system.

To begin our description of the problem, suppose we initialize the system in the $\mathbb{Z}_2$ state which can be accurately expressed  as $\ket{\Psi(0)}=\sum_{l\in \Lambda_s}\psi_l \ket{\phi^{(0)}_l},$ where $\Lambda_s$ is the set of indices of optimized scar states.  We express the time-evolved state in the basis of clean eigenstates in the form $\ket{\Psi(t)}=\sum_{l}\mathcal{A}_{l}(t)e^{-iE^{(0)}_lt}\ket{\phi^{(0)}_l},$ where the sum  runs over the full set of energy eigenstates of the clean system since disorder will inevitably induce transitions to states outside of the scar subspace. To satisfy the initial conditions, we must have that $\mathcal{A}_{l\in\Lambda_s}(0)=\psi_l,$ and $\mathcal{A}_{l\not\in\Lambda_s}(0)=0.$  The disorder-averaged magnetization is then 
\begin{equation}
[\widetilde{\mathcal{M}}_a(\pi,t)]=\sum_{ll'}e^{i\omega_{l'l}t}[\mathcal{A}^*_{l'}(t)\mathcal{A}_{l}(t)]\bra{\phi^{(0)}_{l'}}\widetilde{\sigma}^a_{\pi}\ket{\phi^{(0)}_l},\label{Eq_FullM}
\end{equation}
The magnetization is thus determined by the dynamics of $[\mathcal{A}^*_{l'}(t)\mathcal{A}_{l}(t)]$ as well as the structure of the matrix elements $\bra{\phi^{(0)}_{l'}}\widetilde{\sigma}^a_{\pi}\ket{\phi^{(0)}_l}.$ Any deviations from the clean limit will arise from the factor $[\mathcal{A}^*_{l'}(t)\mathcal{A}_{l}(t)]$ in Eq.~(\ref{Eq_FullM}). This factor can change the oscillation amplitude as well as the oscillation frequency of the magnetization $[\widetilde{\mathcal{M}}_a(\pi,t)].$

\subsection{Decay of optimized scar amplitudes}

Insight into the behavior of the amplitudes $\mathcal{A}_{l\in\Lambda_s}(t)$ can be gained by calculating their dynamics in the weak-disorder limit $W\ll\omega_{\text{scar}}.$   When the disorder strength is weak, scar eigenstates primarily hybridize with nearby states within a spectral neighbourhood of size $\Delta\ll\omega_{\text{scar}},$ as we schematically represent in Fig.\ref{Fig_DecayPic}a.  Since the number of states in this spectral neighbourhood is exponentially large in the system size, the coupling of the scar state to its spectral surrounding leads to the irreversible decay of the amplitude $\mathcal{A}_{l\in \Lambda_s}(t).$ The dynamics corresponds to the decay of a discrete state that is embedded in a quasi-continuum. Standard perturbative calculations of this type of decay \cite{Tannoudji} lead us to derive the approximate expression for the amplitude (see Appendix \ref{App_AA} for details):
\begin{equation}
    \mathcal{A}_{l\in \Lambda_{s}}(t)\approx\psi_{l} e^{-i  \delta\mathcal{E}_l(t)t}e^{-\frac{1}{2}\lambda_l(t) t},\label{Eq_A0A0}
\end{equation} 
where we defined 
\begin{subequations}
\begin{eqnarray}
\delta\mathcal{E}_{l}(t)&=&\bra{\phi^{(0)}_{l}}\hat{W}\ket{\phi^{(0)}_{l}}\label{Eq_deltaE}\\
&&+\sum_{\substack{l'\not\in\Lambda_s\\ \vert\omega_{ll'}\vert<\Delta}}\frac{\vert\bra{\phi^{(0)}_{l}}\hat{W}\ket{\phi^{(0)}_{l'}}\vert^2}{\omega_{ll'}} \left(1-\frac{\sin\omega_{ll'}t}{\omega_{ll'}t}\right),\nonumber\\
\lambda_l(t)&=&2\sum_{\substack{l'\not\in\Lambda_s\\ \vert\omega_{ll'}\vert<\Delta}}\vert\bra{\phi^{(0)}_{l}}\hat{W}\ket{\phi^{(0)}_{l'}}\vert^2 \left(\frac{1-\cos\omega_{ll'}t}{\omega^2_{ll'}t}\right).\label{Eq_lambda}
\end{eqnarray}\label{Eq_dE}
\end{subequations}
The expression Eq.~(\ref{Eq_A0A0}) represents the dynamics induced by a general disorder operator $\hat{W},$ although below we will use the explicit form defined in Eq.~(\ref{Eq_Dis}).  The factor $e^{-i  \delta\mathcal{E}_l(t)t}$ has fixed magnitude and encodes perturbative shifts in the scar energy due to the disorder potential. By contrast, the factor $e^{-\frac{1}{2}\lambda_l(t)t}$ leads to the decay of the scar amplitude. In fact, at sufficiently long times, the factor $(1-\cos \omega_{ll'}t )/(\omega^2_{ll'}t)$ is highly peaked at low frequencies and converges to $\sim\pi  \delta(\omega_{ll'}).$ The sum over states in Eq.~(\ref{Eq_lambda}) can then be turned into an integral that effectively yields the decay rate of the scar state as determined by Fermi's Golden rule. We will not take this limit here since we are dealing with finite sized systems with discrete spectra and, as we shall see, the expressions in terms of discrete sums account very well for the observed dynamics. 

The next step is to calculate the average $[\mathcal{A}^*_{l'\in\Lambda_s}(t)\mathcal{A}_{l\in\Lambda_s}(t)]$ over realizations of the random fields $h_a(r)$ in Eq.~(\ref{Eq_Dis}). To perform this average explicitly, we make use of a Gaussian probability distribution with the same first and second moments, $[h_a(r)]=0$ and $[h^2_a(r)]=W^2/12$ respectively, of the box disorder distribution we used numerically. We are thus led to the explicit disorder-averaged expression
\begin{eqnarray}
    \left[\mathcal{A}^*_{l'\in\Lambda_s}(t)\mathcal{A}_{l\in\Lambda_s}(t)\right]&\approx&\psi^*_{l'}\psi_{l}\sqrt{\det\left\{\alpha_{l'l}(t)\right\}}\label{Eq_A0A0_ave}\\
    &&\,\times e^{-\frac{W^2t^2}{24}\left(\rho_{l'l'}-\rho_{ll}\right)\cdot \alpha_{l'l}(t) \cdot \left(\rho_{l'l'}-\rho_{ll}\right)},\nonumber
\end{eqnarray}
where we defined $(\rho_{ll'})_{n}=\bra{\phi^{(0)}_l}\sigma^{a(n)}_{r(n)}\ket{\phi^{(0)}_{l'}},$ $\mathbf{h}_n=h_{a(n)}(r(n))$ with $ n=1, \ldots, 3L$,  $a(n)=(n-1)\text{mod} 3$,   $r(n)=\left\lfloor \frac{n-1}{3}\right\rfloor,$  we made the identification $\sigma^{1,2,3}_r=\sigma^{x,y,z}_r,$ and \begin{eqnarray}
\alpha^{-1}_{l'l}(t)&=&\mathbb{I}+\frac{W^2}{6}\left\{G^{\dagger}_{l'}(t,0)+G_{l}(t,0)\right\},\\
G_{l}(t_1,t_2)&=&\sum_{\substack{l'\neq l,l'\not\in\Lambda_s\\ \vert\omega_{ll'}\vert<\Delta}}\left(\rho^*_{ll'}\cdot\rho^{T}_{ll'}\right) f_{mn}(t_1,t_2),\label{Eq_Gt1t2}
\end{eqnarray}
where $f_{mn}(t,t')=i\omega^{-2}_{mn}(\omega_{mn}(t-t')-i(\exp{(i\omega_{mn}t')}-\exp{(i\omega_{mn}t)})).$ We defined this function in terms of two temporal arguments for convenience, as it will arise again in this form in the following section. For the same reason, although the sum in Eq.~(\ref{Eq_Gt1t2}) has the redundant constraints $l'\neq l$ and $l'\not\in\Lambda_s,$ it will be useful later when we discuss the amplitudes $\mathcal{A}_{l\not\in\Lambda_s}(t)$

The expression Eq.~(\ref{Eq_A0A0_ave}) shows us that the magnetization decays on account of both the square root and exponential factors, encoding two sources of decay. The exponential factor represents dephasing induced by random energy shifts from Eq.~(\ref{Eq_deltaE}) which are linear in the disorder strength. The factor $\sqrt{\text{det}\{\alpha_{l'l}(t)\}}$ arises in part from the exponential factor in Eq.\ref{Eq_A0A0}, which  encodes information about the finite life time of the scar state as it decays into its spectral neighbourhood.

On the other hand, the magnetization oscillation frequency can also change if $\left[\mathcal{A}^*_{l'\in\Lambda_s}(t)\mathcal{A}_{l\in\Lambda_s}(t)\right]$ develops a time-dependent imaginary component. Inspection of Eq.~(\ref{Eq_A0A0_ave}) reveals that this can happen via the second-order energy shift in Eq.~(\ref{Eq_deltaE}). Since this shift is suppressed with respect to the first order term in the energy shift, we expect deviations from $\omega_{\text{scar}}$ to be small. This explains why, for the disorder strengths in Fig.\ref{Fig_PXP_ParamVsPXP}a, which are small with respect to $\omega_{\text{scar}},$ there appears to be some level of rigidity in the frequency of oscillation of the magnetization. 

\begin{figure}
\centering
 \includegraphics[scale=0.23]{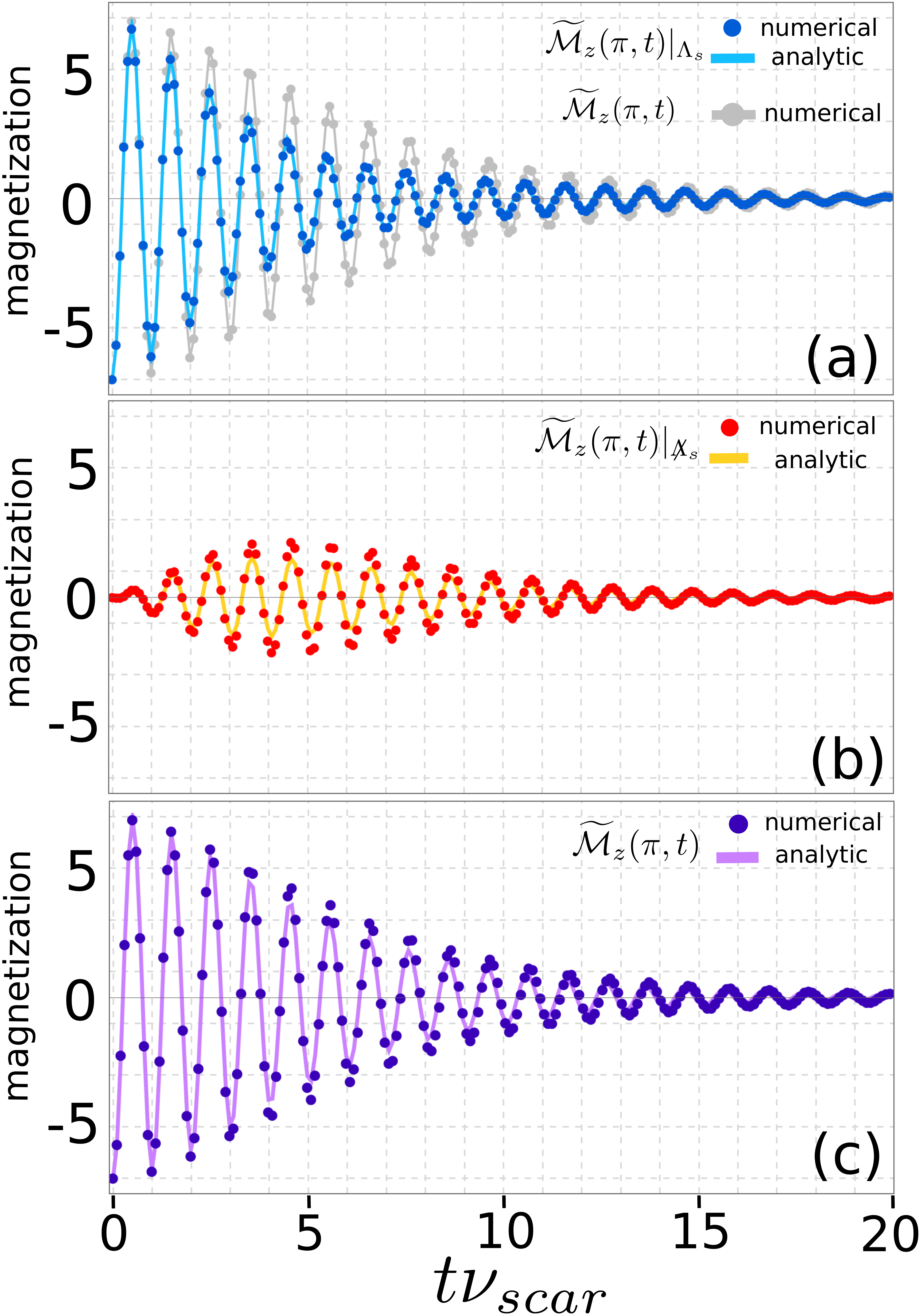}
\caption{Comparison of contributions to the magnetization $[\widetilde{\mathcal{M}}_z(\pi,t)]$ when evaluated exactly numerically (dots) and analytically (solid lines) for $L=14$ and $W=0.1\,\omega_{\text{scar}}.$ The matrix elements $\rho_{ll'}$ were calculated numerically using the eigenstates of the clean system. \textbf{(a)} Contribution from optimized scar states, $J = 1$ tower, with the analytic curve obtained using  Eq.~\ref{Eq_MJ1}.  \textbf{(b)} Contribution from states outside the main scar subspace, decay via $J\ne 1$ towers as shown in \ref{Fig_DecayPic},  with the corresponding analytic curve obtained from Eq.~(\ref{Eq_MJRest}). \textbf{(c)} Contribution from all states in the spectrum, with the corresponding analytic curve. In \textbf{(a)}, we also show the net magnetization by gray dots (the connecting gray line is a guide for the eye) to illustrate that the contribution from the optimized scar states is not sufficient to account for the observed dynamics.}\label{Fig_MagnetizationComparison}
\end{figure}

We now use Eq.~(\ref{Eq_A0A0_ave})  to compute the contribution of the scar states to the magnetization using the restricted sum
\begin{equation}
    [\widetilde{\mathcal{M}}_a(\pi,t)]\rvert_{\Lambda_s}=\sum^{L}_{m=1}2\text{Re}\left(e^{i\Omega_{1m}t}\mathcal{Q}_{1m}(t)\Gamma^{a,+1}_{1m}\right),\label{Eq_MJ1}
\end{equation}
where we defined $\mathcal{Q}_{Jm}(t)=[\mathcal{A}^*_{Jm+1}(t)\mathcal{A}_{Jm}(t)],$ and $\Omega_{Jm}=\omega_{Jm+1,Jm}.$ In Fig.\ref{Fig_MagnetizationComparison}a we present $[\widetilde{\mathcal{M}}_a(\pi,t)]\rvert_{\Lambda_s}$ for the numerically-obtained values and our explicit disorder-averaged result. In calculating the analytic result, the matrix elements $\rho_{ll'}$ were calculated numerically using the eigenstates of the clean system. Both match very closely, which confirms that our basic picture of the decay of scar amplitudes is accurate. When compared, however, with the full magnetization signal, this contribution is not sufficient to account for the full magnetization of the system, even though it clearly is sufficient in the clean limit. The inevitable conclusion is that states \textit{outside} of the scar subspace are contributing to the dynamics of the system. What is perhaps most remarkable is that this additional contribution must oscillate at the  same frequency $\omega_{\text{scar}},$ even though it involves states that are known to be more ergodic and more highly entangled than the optimized scar states. This requires us to investigate how such states can coherently contribute to the observed magnetization, as we do so in the following section.

\subsection{Decay of $J \neq 1$ scar amplitudes}

The key to identifying the missing contribution to the magnetization can be gleaned from Eq.~(\ref{Eq_FullM}). It is clear that as the scar states become de-populated, amplitudes for states outside the scar subspace will inevitably begin to grow. In order to obtain the dynamics arising from the other towers, the decay of the amplitudes $\mathcal{A}_{l \not \in \Lambda_s}(t)$ can be calculated similar to the case of optimized scars, as we schematically represent in Fig.\ref{Fig_DecayPic}b.  Following the same main steps we took for the set of optimized scars, the solution for $\mathcal{A}_{l \not \in \Lambda_s}(t)$ can be written in the integral form (see Appendix \ref{App_AA} for details)
\begin{eqnarray}
    \mathcal{A}_{l \not \in \Lambda_s}(t)&\approx&-i \bra{\phi^{(0)}_{l}}\hat{W}\ket{\phi^{(0)}_{n_0(l)}}\int^{t}_0e^{it\omega_{ln_0(l)}}\mathcal{A}_{n_0(l)}(\tau)\label{Eq_AJm0}\\
    && \qquad\qquad\times e^{-i  \delta\mathcal{E}_l(t,\tau)(t-\tau)}e^{-\frac{1}{2}\lambda_l(t,\tau) (t-\tau)}d\tau\nonumber,\label{Eq_AnE}
\end{eqnarray}
where we defined
\begin{eqnarray}
&&\delta\mathcal{E}_{l}(t,\tau)=\bra{\phi^{(0)}_{l}}\hat{W}\ket{\phi^{(0)}_{l}}\nonumber\\
&&\quad \, +\sum_{\substack{l'\neq l,l'\not\in\Lambda_s\\ \vert\omega_{ll'}\vert<\Delta}}\frac{\vert\bra{\phi^{(0)}_{l}}\hat{W}\ket{\phi^{(0)}_{l'}}\vert^2}{\omega_{ll'}}\left(1-\frac{\sin\omega_{ll'}t-\sin\omega_{ll'}\tau}{\omega_{ll'}(t-\tau)}\right),\nonumber\\
&&\lambda_l(t,\tau)=2\sum_{\substack{l'\neq l,l'\not\in\Lambda_s\\ \vert\omega_{ll'}\vert<\Delta}}\vert\bra{\phi^{(0)}_{l}}\hat{W}\ket{\phi^{(0)}_{l'}}\vert^2 \left(\frac{\cos\omega_{ll'}t-\cos\omega_{ll'}\tau}{\omega^2_{ll'}(t-\tau)}\right).\nonumber
\end{eqnarray}
Here, $n_0(l)$ refers to the optimized scar state that is spectrally closest to the $l$-th state, as illustrated in Fig.\ref{Fig_DecayPic}b. These functions are straightforward generalizations of $\delta \mathcal{E}_l(t)$ and $\lambda_l(t)$ in Eq.~(\ref{Eq_dE}).
To get an intuitive sense of how the solution Eq.~(\ref{Eq_AnE}) behaves, we can momentarily take $\delta \mathcal{E}_l(t,\tau)\rightarrow \delta \overline{\mathcal{E}}_l$ and $\lambda_l(t,\tau)\rightarrow \overline{\lambda}_l$ to be constant since they are expected to be slowly varying. We similarly do this for $\delta \mathcal{E}_{n_0(l)}(t)\rightarrow \delta \overline{\mathcal{E}}_{n_0(l)}$ and $\lambda_{n_0(l)}(t)\rightarrow \overline{\lambda}_{n_0(l)}.$ We can then perform the time integral in Eq.~(\ref{Eq_AnE}) explicitly to obtain
\begin{eqnarray}
    \mathcal{A}_{l\not\in\Lambda_s}(t)&\propto& \exp\left\{-i  \delta\overline{\mathcal{E}}_l t-\frac{1}{2}\overline{\lambda}_l t\right\}\\
    && -\exp\left\{-i  (\delta\overline{\mathcal{E}}_{n_0(l)}-\omega_{ln_0(l)})t-\frac{1}{2}\overline{\lambda}_{n_0(l)} t\right\}.\nonumber
\end{eqnarray}
This expression illustrates the overall behavior of the amplitudes $\mathcal{A}_{l\not\in\Lambda_s}(t).$  Its absolute value vanishes at $t=0$ as well as at long times, reaching a maximum at intermediate times.  The time scales in this evolution are the decay rate $\overline{\lambda}_{n_0(l)}$ of the optimized scar states that play the role of sources for the $J\neq 1$ towers; and the decay rate $\overline{\lambda}_{l}$ of the $J\neq 1$ towers themselves which transition further to their spectral neighbourhood. The balance between these two time scales determines the maximum value reached by $\vert\mathcal{A}_{l\not\in\Lambda_s}(t)\vert.$ Note from Eq~(\ref{Eq_AnE}) that this maximum value is also controlled by the random matrix element $\bra{\phi^{(0)}_{l}}\hat{W}\ket{\phi^{(0)}_{n_0(l)}}.$

The final step we need to take is to perform the disorder average  $\left[\mathcal{A}^*_{l'\not\in\Lambda_s}(t)\mathcal{A}_{l\not\in\Lambda_s}(t)\right].$ Since it is somewhat more complicated and not very illuminating, we leave its explicit expression in the Appendix \ref{App_AA}. Using this expression, we can then calculate the contribution to the magnetization from the $J\neq 1$ towers as 
\begin{equation}
    [\widetilde{\mathcal{M}}_a(\pi,t)]\rvert_{\not\Lambda_s}\approx\sum_{J\neq 1}\sum^{\mathcal{D}_J-1}_{m=1}2\text{Re}\left(e^{i\Omega_{Jm}t}\mathcal{Q}_{Jm}(t)\Gamma^{a,+1}_{Jm}\right).\label{Eq_MJRest}
\end{equation}
In Fig.\ref{Fig_MagnetizationComparison}b, we compare our explicit expression (red dots) with the numerically-obtained contribution from the lower towers (yellow solid line), finding good agreement between them. The close match confirms our assumptions about the existence of lower towers as well as the manner in which they become populated. In Fig.\ref{Fig_MagnetizationComparison}c we add the analytic result when summed over all towers and again find good agreement with the numerical values for the full magnetization. Although it is more subtle in this case to infer that the $J\neq 1$ towers do not change the oscillation frequency, in Fig.\ref{Fig_MagnetizationComparison}c we see that the magnetization from lower towers continues to oscillate at a frequency close to the scar frequency.

\begin{figure}
\centering
 \includegraphics[trim = 0mm 0.cm 0cm 0mm, clip,scale=0.12]{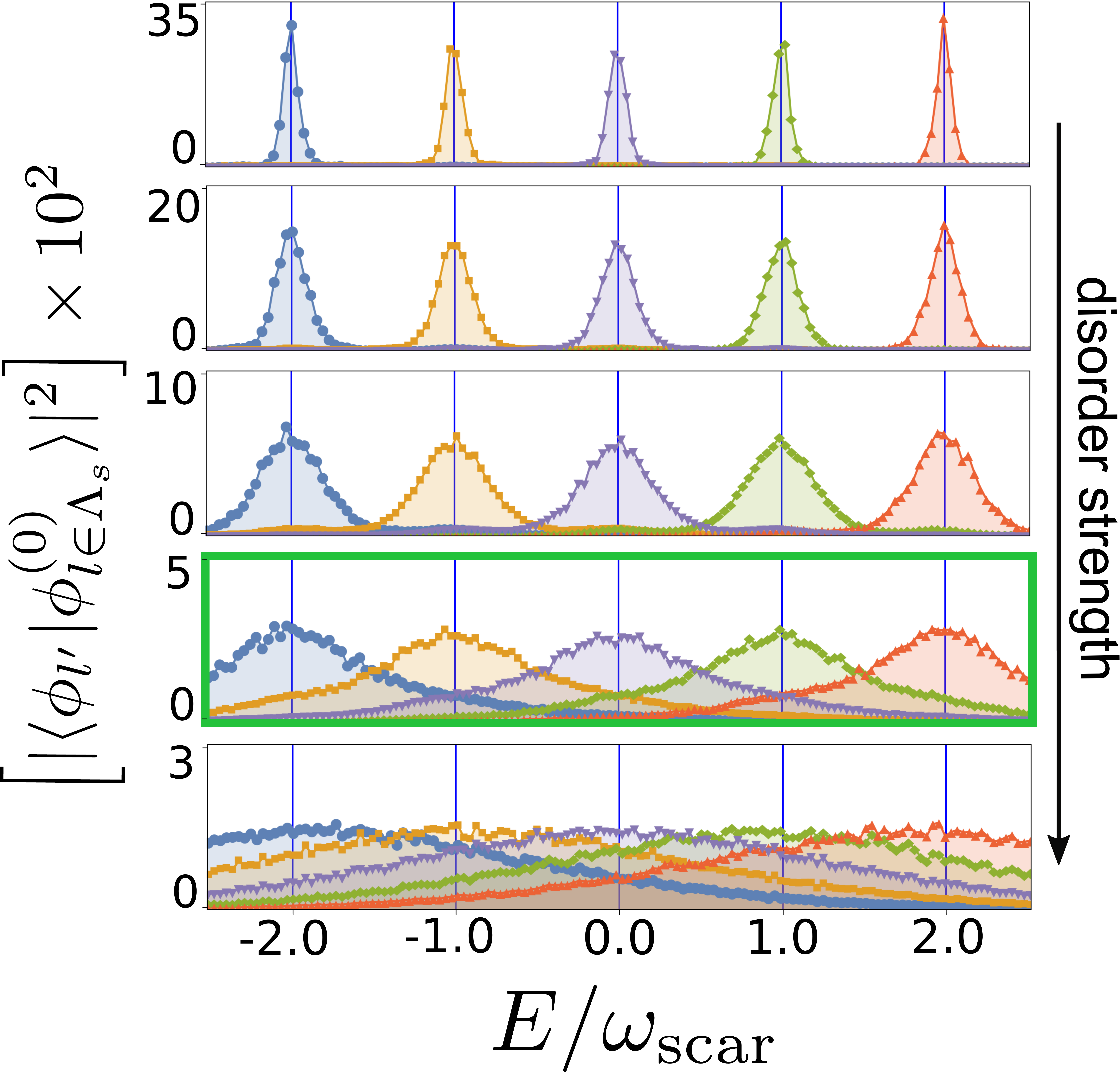}
\caption{Scar resonances in the spectrum. We plot the  distribution of optimized scar states in the basis of disordered eigenstates for $L=12$ and disorder strengths for $W/\omega_{\text{scar}}=0.10, 0.19, 0.34, 0.63, 0.17$ from top to bottom, respectively. To make it easier to visualize the shape of these distributions, we binned the energy axis and summed the probabilities inside each bin. As can be seen, for sufficiently weak disorder, the scar states are broadened in the disordered setting but remain distinguishable. Eventually, at a disorder strength $W_c\approx 0.63 \omega_{\text{scar}}$ (indicated here by the green frame), the width of these states increases so much that they can no longer be spectrally resolved.}\label{Fig_resonances}
\end{figure}

\subsection{Distribution and stability of scar resonances}

Our description of the disordered system suggests thinking about the PXP spectrum as being comprised of multiple towers of scars states that exhibit varying levels of ergodicity and entanglement. Upon disordering the system, these approximate scar states acquire a finite life time, making them effectively resonances in the disordered spectrum. To illustrate the formation of scar resonances in the system, we can calculate the distribution of the clean scar states in the basis of disordered eigenstates. In Fig.\ref{Fig_resonances}, we show the disorder-averaged and coarse-grained probability $[\vert \langle\phi_l\vert\phi^{(0)}_{l\in\Lambda_s}\rangle\vert^2]$ as a function of the energy $E_l$ of the disordered eigenstate $\ket{\phi_l}.$ To make it easier to visualize the overall shape of these distributions,  we binned $\vert \langle\phi_l\vert\phi^{(0)}_{l\in\Lambda_s}\rangle\vert^2$ in discrete ranges of energy and summed them over each bin. For weak disorder relative to $W_c,$ clear spectrally-localized distributions are obtained that are centered at each of the scar energies of the clean limit. As the disorder strength is increased, their widths increase, which is what leads to the decay of the magnetization. Since their width is less than $\omega_{\text{scar}},$ the distribution corresponding to each scar state is clearly distinguished from each other. As long as the width of these resonances is small compared to the scar frequency, oscillations at the scar frequency will continue to be resolved in the dynamics. Importantly, the distributions remain centered at the scar energies, which is consistent with the rigidity in the oscillation frequency we found in Fig.\ref{Fig_PXP_ParamVsPXP}a, which we also concluded from our analytic result Eq.~(\ref{Eq_A0A0_ave}).  When the disorder strength becomes sufficiently strong, the distributions overlap significantly (as illustrated by the green box in Fig.\ref{Fig_resonances}), and thus we expect oscillations at the scar frequency to disappear.

This picture of resonances allows us to infer an approximate value for $W_c.$ Given that the contribution from $J\ne 1$ towers is smaller and appears at finite times, we focus on the main scar tower to determine this value. The width of these distributions in the frequency domain determines the decay of the scar amplitudes $\mathcal{A}_{l\in\Lambda_s}(t).$ As long as this broadening is sufficiently small, $\mathcal{A}_{l\in\Lambda_s}(t)$ will evolve slowly with respect to the scar period $2\pi/\omega_{\text{scar}}.$  When the disorder strength is increased, their widths continue to grow, up until when $W$ approaches $W_c,$ at which point the distributions overlap significantly and are no longer distinguishable. We expect that $W_c$ is the disorder strength for which, after a time $T=2\pi/\omega_{\text{scar}},$ the probability $\vert\mathcal{A}_{l\in \Lambda_s}(T)\vert^2=e^{-\lambda_l(T)T}$ is reduced to $1/e$ of its value at $t=0.$ The disorder averaged result, obtained from Eq.~(\ref{Eq_A0A0_ave}), leads to
\begin{equation}
   \frac{1}{\sqrt{\det\left(\mathbb{I}+\frac{W_c^2}{6}\left\{G^{\dagger}_{l\in \Lambda_s}(T,0)+G_{l\in \Lambda_s}(T,0)\right\}\right)}}\sim e^{-1}.
\end{equation}
This produces the value $W_{c}\approx 0.34 \omega_{\text{scar}}$ for scar states near the middle of the spectrum for $L=14,$ which is of the order of the value $W_c\approx 0.63 \omega_{\text{scar}}$ we found numerically in Fig.\ref{Fig_PXP_flow}b.

There is an important and subtle question regarding the scaling to the thermodynamic limit. Even though we found numerically that the magnetization of the system did not change appreciably with system size (see Fig.\ref{Fig_PXP_flow}a), it was argued in \cite{Lin2020} that perturbations that are added to a system with optimized scar states must inevitably thermalize for a sufficiently large system. Furthermore, using Lieb-Robinson bounds \cite{Lin2020}, it was argued that it would still be possible to observe non-ergodic physics for a time scale $t^*\sim \mathcal{O}\left(\epsilon^{-\frac{1}{2}}\right),$ where $\epsilon$ is the strength of the perturbation. While in \cite{Lin2020} the PXP Hamiltonian and the perturbation were clean systems, their argument has some bearing on our results. The reason that scars might be expected to thermalize in the disordered system is that matrix elements of the disorder operator between a scar state and another state that is ergodic must scale as $\mathcal{D}^{-1/2},$ whereas the density of states scales as $\mathcal{D}$ ($\mathcal{D}$ being the dimension of the Hilbert space). This suggests that the factor $\lambda_l(t)$ in Eq.~(\ref{Eq_lambda}) can potentially remain finite, so that non-ergodic dynamics occurs for a time scale $t^*\sim \mathcal{O}\left(W^{-2}\right),$ consistent with \cite{Lin2020}. However, higher order terms in the perturbative calculation of the amplitude might get overwhelmed by the exponential  growth of the density of states. Although we do not find numerical evidence of this in this work, we cannot rule out this possibility. We do note, however, that the arguments in \cite{Lin2020} assume that all states outside of the scar subspace are exactly ergodic. Instead, as we have found here, they realize additional scar towers which, although more highly entangled, are not completely volume-law states. This adds another subtle layer to the problem, which warrants further investigation.

\begin{figure}
\centering
 \includegraphics[trim = 0mm 0.cm 0cm 0mm, clip,scale=0.18]{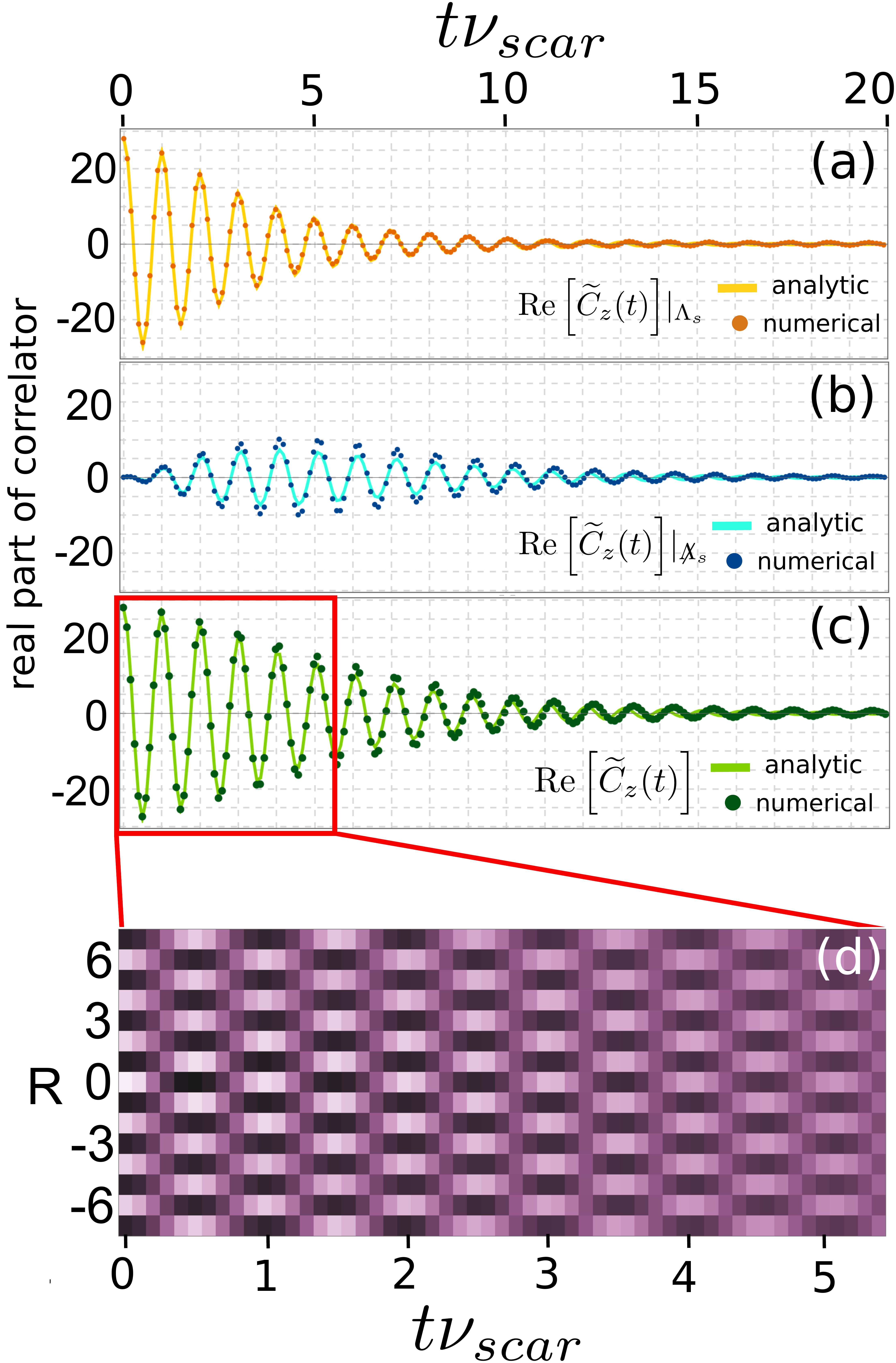}
\caption{ Spatio-temporal correlations  for $L=14$ and $W=0.1\,\omega_{\text{scar}}$ starting from  clean scar eigenstate at zero energy as the initial state. The plots in \textbf{(a,b,c)} correspond to the different contributions to the net temporal correlator $\text{Re}\left[\widetilde{C}_z(t)\right].$  The solid lines correspond to the analytic result Eq.~(\ref{Eq_TCorr}) whereas the dots are numerically exact values. \textbf{(d)} Spatial resolution of the correlator $[\vert \mathcal{C}_z(r_0,r_0 +R,t)\vert]$ (with $r_0$ an odd site) that underlies the oscillations found in the temporal correlator.}\label{Fig_ScarCorrelator}
\end{figure}

\section{Spatio-temporal correlations}\label{Sec_Correlations}

Our quantitative perturbative description of the magnetization dynamics can also be extended to provide physical understanding of spatio-temporal correlations starting from different initial states, such as clean scar eigenstates or the $\mathbb{Z}_2$ product state. It was shown in Ref. \cite{Iadecola2019} that the equal-time connected correlator $\bra{\phi^{(0)}_n} \left(\mathcal{P}\sigma^z_{r}(0)\mathcal{P}\right)\left(\mathcal{P}\sigma^z_{r+R}(0) \mathcal{P}\right)\ket{\phi^{(0)}_n}_{c}$ evaluated with respect to an optimized scar eigenstate converges to a nonzero value for large $R,$ suggesting that the states are long-range spatially ordered. Furthermore, it can be argued that scars are temporally ordered as well \cite{Iadecola2019}. We here consider the spatio-temporal correlator
\begin{equation}
\mathcal{C}_a(r_0,R,t)=\bra{\Psi} \left(\mathcal{P}\sigma^a_{r_0}(t)\mathcal{P}\right)\left(\mathcal{P}\sigma^a_{r_0+R}(0)\mathcal{P}\right) \ket{\Psi}_{c}.\label{Eq_correlator}
\end{equation}
Temporal correlations can be explicitly calculated for the spatial Fourier transform  $\widetilde{\mathcal{C}}_a(t)=\sum^L_{R,r_0=1}\int dt e^{i\pi R}\mathcal{C}_a(r_0,r_0+R,t).$ Following the same steps as in the calculation of the magnetization, we find that (see Appendix \ref{App_BB} for details):
\begin{equation}
[\widetilde{\mathcal{C}}_a(t)]=\mathcal{T}_a(t)-[\widetilde{\mathcal{M}}_a(\pi,t)]\widetilde{\mathcal{M}}_a(\pi,0),\label{Eq_TCorr}
\end{equation}
where
\begin{eqnarray}
\mathcal{T}_a(t)&\approx&\sum_{Jm}\sum_{sd}\left(\frac{\overline{\psi}^a_{n_0(Jm-sd)}\Gamma^{a,sd}_{Jm-sd}}{\psi_{n_0(Jm)}}\right)\\
&&\qquad\qquad\quad \times[\mathcal{A}^*_{Jm+s d}(t)\mathcal{A}_{Jm}(t)]e^{it\omega_{Jm+sd,Jm}}. \nonumber
\end{eqnarray}
In this expression, we defined $\overline{\psi}^a_{m}=\sum_{s,d}\psi_{J=1,m-sd}\Gamma^{a,sd}_{1,m-s d}$ with $s=\pm 1,$ $d=1,3.$ For convenience, in these sums we set $\psi_{J,m}=\Gamma^{a,sd}_{J,m}=0$ whenever $m$ is out of range in the sum. In contrast to the magnetization, for the calculation of temporal correlators we need to include matrix elements beyond the tri-diagonal that are small but nevertheless contribute to the final temporal correlator. The expression Eq.~(\ref{Eq_TCorr}) is completely determined, since we have already calculated  $[\mathcal{A}^*_{l'}(t)\mathcal{A}_{l}(t)]$ in previous sections. Similar to the magnetization, the dynamics of temporal correlators is thus also controlled by the existence of multi-tower scar resonances. 

The expression Eq.~(\ref{Eq_TCorr}) depends on both the choice of axis index $a$ of the Pauli operators, and the initial state $\ket{\Psi}.$ As a first example, we calculate the correlator using the clean scar eigenstate at ${E=0}$ for magnetization component along $a=z$ axis. In Fig.\ref{Fig_ScarCorrelator} we show the contributions from the $J=1$ and $J\neq 1$ towers for both the numerical (dots) and analytic (solid lines) values, showing again very good agreement between both. The corresponding spatio-temporal correlator $[\vert \mathcal{C}_z(r_0,r_0+R,t) \vert]$ in the $(R,t)$ plane (with $r_0$ an odd site) underlying these results is shown in Fig.\ref{Fig_ScarCorrelator}a. The zero-energy scar state clearly leads to a periodic pattern in the $(R,t)$ plane indicating that spatially separated qubits are periodically correlated in time.  While correlations are reduced by disorder as the temporal separation is increased, we have not observed  decay as a function of $R.$ This is counter intuitive, as disorder usually leads to a decay of spatial correlations with a characteristic length scale determined by the disorder strength. In this case, this decay appears to be absent, suggesting that long-range order is maintained. This is quite remarkable, and deserves more detailed study as a function of system size and disorder strength. It is also reminiscent of spatio-temporal correlations in time crystals, the  non-equilibrium phases of matter that spontaneously break the time-translational symmetry of the Hamiltonian \cite{Khemani2016,Else2016,Zhang2017}. Time-translation symmetry is broken when  $\lim_{R\rightarrow\infty}\mathcal{C}_a(r,R,t)$ oscillates with a frequency that differs from the temporal period of the Hamiltonian and is robust to small perturbations, signifying rigidity \cite{Yao2017}. In the scar case, the Hamiltonian has full (in contrast to discrete) time translation symmetry, and we indeed find evidence of spatial long-range order with a rigid frequency of oscillation that remains close to the clean scar frequency in spite of the presence of disorder.

\begin{figure}
\centering
 \includegraphics[trim = 0mm 0.cm 0cm 0mm, clip,scale=0.18]{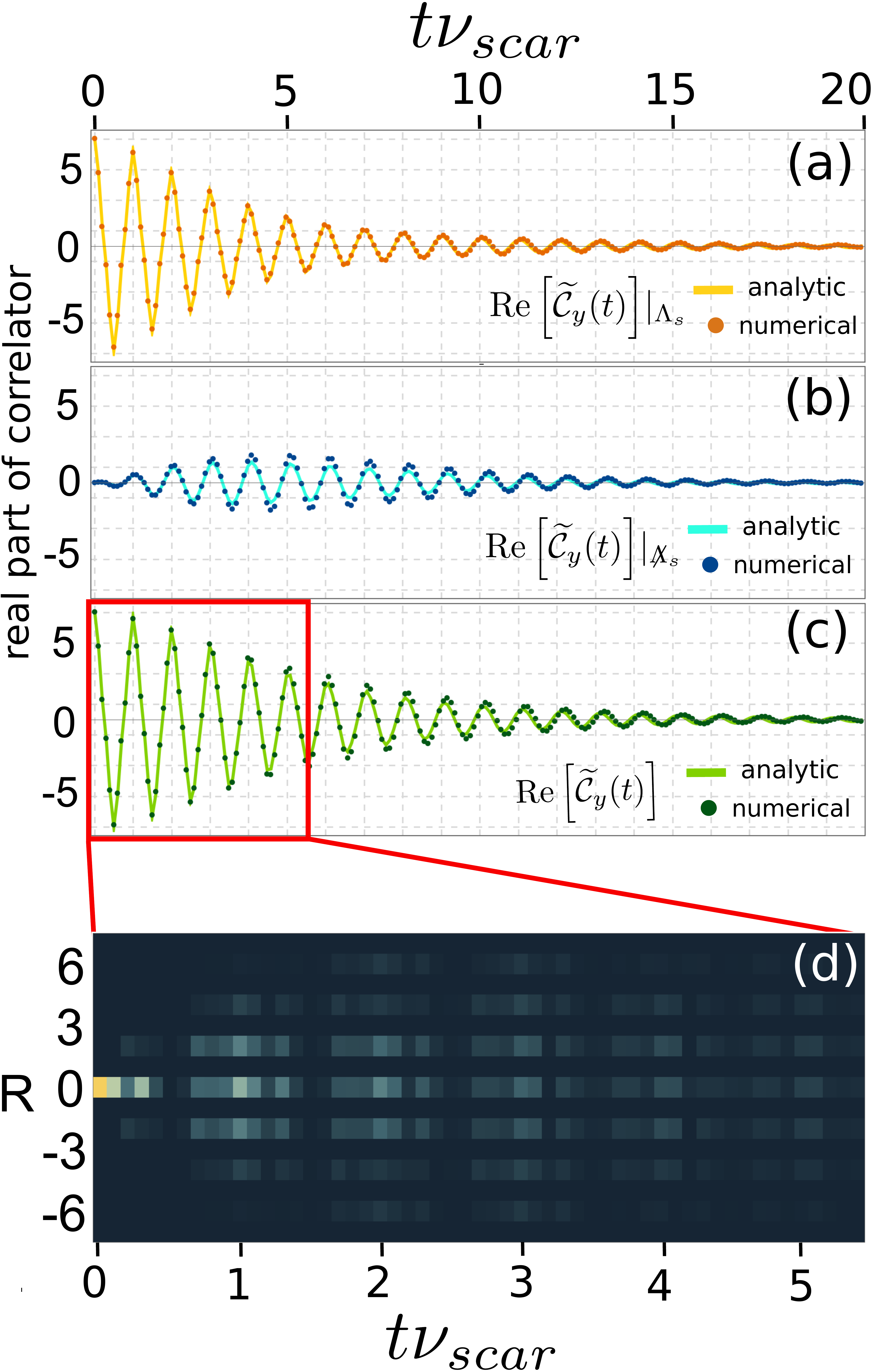}
\caption{Spatio-temporal correlations  for $L=14$ and $W=0.1\,\omega_{\text{scar}}$ starting from  the $\ket{\Psi}=\ket{\mathbb{Z}_2}$ state. The plots in \textbf{(a,b,c)} correspond to the different contributions to the temporal correlator $\text{Re}\left[\widetilde{C}_y(t)\right].$ The solid lines correspond to the analytic result Eq.~(\ref{Eq_TCorr}) whereas the dots are numerically exact values.  \textbf{(d)} Spatial resolution of the correlator $[\vert \mathcal{C}_y(r_0,r_0 +R,t)\vert]$ (with $r_0$ an odd site) that underlies the oscillations found in the temporal correlator.}\label{Fig_Z2correlator}
\end{figure}

While evaluating correlators starting with the scar eigenstates is illuminating, these states may not be  easily  accessible experimentally.  For this reason, we also consider evaluating correlations starting from the $\ket{\Psi}=\ket{\mathbb{Z}_2}$ initial state. In this case, connected correlators with respect to the $\sigma^z_r$ operators vanish, so instead we analyze the $a=y$ component. In Fig.\ref{Fig_Z2correlator} we again show the contributions from the $J=1$ and $J\neq 1$ towers for both the numerical (dots) and analytic (solid lines) values, showing again very good agreement between both. Notably, although oscillations continue to occur at the scar frequency, the underlying spatio-temporal pattern is very different from the pattern that we obtained starting from a scar eigenstate. Now the correlation function begins concentrated at $R=0,$ and  spreads out as a function of time until it covers the full extent of the system. These correlations can in principle be probed in experiments. However, note that these results involve projected Pauli operators. To make even closer connection with future experiments, in the following section we evaluate them for full Rydberg Hamiltonian without any projection operators.

\begin{figure*}
\centering
 \includegraphics[trim = 0mm 0.cm 0cm 0mm, clip,scale=0.4]{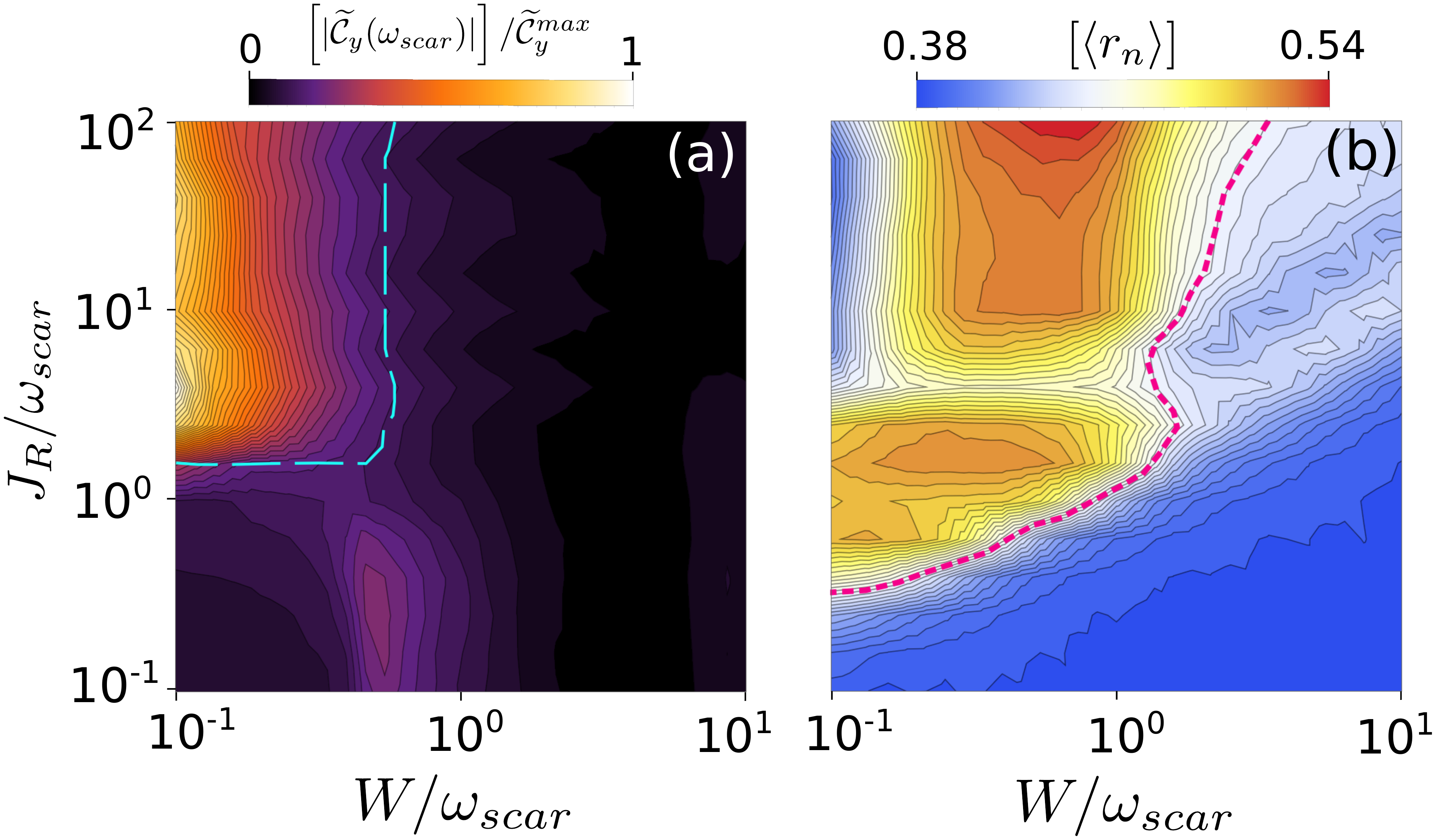}
\caption{\textbf{(a)} Contour plot of  $[\vert\widetilde{\mathcal{C}}^{Ryd}_y(\omega_{\text{scar}})\vert],$ normalized by the maximum value $\widetilde{C}^{max}_y$ in this parameter space.  The dashed line corresponds to the location of the peak in $\overline{\Lambda}^{(q=6)}_{\omega}$ for $L=12.$ \textbf{(b)} Contour plot of the mean energy level spacing ratio $[\langle r_n \rangle]$ that varies between the Poisson value ($[\langle r_n \rangle]=0.38$) and the ergodic value ($[\langle r_n \rangle]=0.6$). The dashed line corresponds to approximately the value $[\langle r_n \rangle]=0.45,$ which we use to estimate the boundary between thermal and the MBL phases. }\label{Fig_JvW_Scar}
\end{figure*}

\section{Diagram of dynamical regimes for the Rydberg Hamiltonian} \label{Sec_PD}

We now return to the full Rydberg Hamiltonian to examine the dynamics arising from scar resonances in the full parameter space $(W, J_R).$ This amounts to removing all projection operators in Eq.~(\ref{Eq_FullModel}). Since the Hilbert space of the Rydberg Hamiltonian grows as $2^L,$ we are constrained in this section to smaller system sizes $L=8,10,12.$  We focus on the connected temporal correlator 
\begin{equation}
[\widetilde{\mathcal{C}}^{Ryd}_y(t)]=[\bra{\mathbb{Z}_2} \widetilde{\sigma}^y_{\pi}(t)\widetilde{\sigma}^y_{\pi}(0)\ket{\mathbb{Z}_2}_{c}].\label{Eq_Rydcorrelator}
\end{equation}
This correlator differs from Eq.~(\ref{Eq_TCorr}) by the absence of projection operators. This makes it a directly experimentally accessible quantity, although it has the downside that we can no longer evaluate it analytically. In Fig.\ref{Fig_JvW_Scar}a, we show a contour plot of its Fourier transform $[\vert\widetilde{\mathcal{C}}^{Ryd}_y(\omega_{\text{scar}})\vert]$ evaluated at the scar frequency in the $(W, J_R)$ parameter space. Oscillations at the scar frequency occur for interaction strengths greater than $J_R\approx \Omega$ and for disorder strengths up to around $W\sim W_c.$ To pinpoint the boundary for this region, we again resort to the frequency participation parameter adapted to this correlation function $\overline{\Lambda}^{(q)}_{\omega}=\left[\int d \omega \mathcal{F}^6(\omega)\right]^{-1},$ where $\mathcal{F}(\omega)=\frac{1}{\mathcal{N}}[\vert\widetilde{\mathcal{C}}^{Ryd}_y(\omega)\vert]$ and $\mathcal{N}=\sqrt{\int d \omega[\vert\widetilde{\mathcal{C}}^{Ryd}_y(\pi,\omega)]\vert^2}.$ We find a line of parameters along which $\overline{\Lambda}^{(q)}_{\omega}$ reaches a maximum, as presented in show Fig.\ref{Fig_JvW_Scar}a with a dashed blue line. This boundary  encompasses a region that is consistent with high values of $[\vert\widetilde{\mathcal{C}}^{Ryd}_y(\omega_{\text{scar}})\vert].$ There is thus an ample regime of parameters for which it is possible to observe non-ergodic dynamics, even for moderate interaction strengths and finite disorder. Interestingly, signatures of scar resonances are present even for low interactions when $J_R\sim \Omega.$

We further examine the ergodicity of the system using the mean level spacing ratio $[\langle r_n\rangle]$. Since the energy spectrum shifts significantly throughout the $(W,J_R)$ parameter space, we average over the spectral range $[\langle H\rangle_{\mathbb{Z}_2}-\delta E_{\mathbb{Z}_2},\langle H\rangle_{\mathbb{Z}_2}+\delta E_{\mathbb{Z}_2}]$ where $\delta E_{\mathbb{Z}_2}=\sqrt{\langle H^2\rangle_{\mathbb{Z}_2}-\langle H\rangle^2_{\mathbb{Z}_2}}$ is the energy spread of the $\mathbb{Z}_2$ state with respect to the Rydberg Hamiltonian, since this is the dynamically relevant part of the spectrum. In Fig.\ref{Fig_JvW_Scar}b, we show a contour plot of $[\langle r_n\rangle]$ in the $(W,J_R)$ plane. A clear region is discernible for which $[\langle r_n\rangle]$ is close to the ergodic value.  A rough boundary for this region can be estimated to occur at the contour  $[\langle r_n\rangle]\sim 0.45,$ which is the approximate value obtained in Fig.\ref{Fig_ergodicity}, and is shown by a magenta dashed in line in Fig.\ref{Fig_JvW_Scar}b. The value of $[\langle r_n\rangle]$ generally increases to the left of this line as the system size is increased through $L=8,10,12,$ and decreases to the right of this line. We note that due to the limited system sizes, there are some parameters points for which it is difficult to clearly discern a direction of flow for $[\langle r_n\rangle]$ as a function of $L.$ Notwithstanding this, the boundary $[\langle r_n\rangle]\sim 0.45$ is consistent at strong interactions with the transition at the disorder strength $W_{Th-L}$ we found for the PXP model.  It is also consistent with the weakly-interacting limit, where $[\langle r_n \rangle]$ indeed is very close to the Poisson value, as expected for an MBL phase which corresponds to a weakly interacting  disordered paramagnet that is expected to be localized. Note that, similar to what we found in Fig.\ref{Fig_ergodicity}, there is a region to the right of the ergodic boundary for which, even though $[\langle r_n\rangle]$ flows to lower values as a function of system size, it decreases very slowly. This is consistent with the constrained MBL phase we found in the PXP limit. 

Overall, it is clear that the ergodic region in Fig.\ref{Fig_JvW_Scar}b is larger than the region of scar resonances in Fig.\ref{Fig_JvW_Scar}a. We thus obtain the overall diagram of dynamical regimes presented in Fig.\ref{Fig_system}a.  There are four regimes as a function of disorder and interaction strengths: a regime with scars resonances, another with a fully ergodic spectrum, a constrained MBL phase, and finally an MBL phase without kinematic constraints. Of note is an unexpected behavior within the ergodic region around $J_R \approx 4 \omega_{\text{scar}},$ where $[\langle r_n\rangle]$ appears to indicate weak ergodicity. Notably, there is a slight improvement in the strength of oscillations at the scar frequency in Fig.\ref{Fig_JvW_Scar}a in this region, perhaps suggesting that the weaker ergodicity is less detrimental for the existence of anomalous dynamics for these interaction strengths.

\section{Discussion and conclusions} \label{Sec_Conclusions}

In this work, we studied the impact of disorder on systems with quantum many-body scars. We found that scar eigenstates become resonances that have a finite life time and remain centered at the scar energies of the clean system. As a result, observables exhibit non-ergodic oscillations at the same scar frequency of the clean limit for a time scale $t^*\sim\mathcal{O}\left(W^{-2}\right).$ We confirmed this picture of scar resonances by calculating explicitly the disorder-averaged magnetization and temporal correlators of the system, which match closely with the values obtained numerically. We further examined the range of interaction and disorder strengths for which scar resonances are present in the Rydberg model, and mapped out a diagram of other dynamical regimes that are accessed when scar resonances are lost at strong disorder and weak interactions.

A natural system where these results can be probed is of course the Rydberg-atom simulator used to originally discover scar states. In this system, scars were observed for $J_R\sim 2\pi \times 24\,$MHz and $\Omega\sim 2\pi \times 2\,$MHz \cite{Bernien2017}. As a result $\omega_{\text{scar}}\sim 2\pi \times 1.3\,$MHz and $J_R\approx 19 \omega_{\text{scar}},$ which places this experimental system well within the regime that has scar eigenstates along the $W=0$ line in Fig.\ref{Fig_system}a. The type of Zeeman disorder we considered can be implemented  using drives with spatially-varying Rabi frequencies. This platform should thus make it possible to observe the changes in dynamical regimes that arise both as a function of disorder as well as interaction strengths.     

Superconducting qubit systems constitute another possible quantum simulator in which to study scar resonances. In charge qubit systems, for example, it is possible to realize the model \cite{Viehmann2013}
\begin{equation}
    H=\sum^L_{r=1}\left(\frac{\Omega_z}{2}\tau^{z}_r+\frac{\Omega_{x}}{2}\tau^x_r+J_{xx}\tau^x_{r}\tau^x_{r+1}\right),\label{Eq_FullIsing}
\end{equation}
which constitutes the transverse Ising model in the presence of a longitudinal field. The $\tau^a_r$ are Pauli matrices that act on the superconducting qubit charge states $\left\{\ket{0},\ket{1}\right\}.$ Up to a constant term and a unitary rotation, this model is equivalent to the Rydberg Hamiltonian Eq.~(\ref{Eq_main}) when  $\Omega_x=4J_{xx}=J_R,$ which is what leads to the kinematic constraint in the system.  In order to reach the regime of scar resonances, according to the diagram in Fig.\ref{Fig_system}a, we must further have $\Omega_z< J_{xx}.$ In these systems, it is possible to reach  $J_{xx}\sim2\pi \times 100\,$MHz, as well as have a flux-tunable qubit frequency $\Omega_z$ \cite{Viehmann2013}, so reaching the scar resonance regime is feasible.  Interestingly, Ising-type models of the form Eq.~(\ref{Eq_FullIsing}) can be realized in trapped ions as well \cite{Monroe2019}. However, an inevitable new ingredient in such a system is the presence of power-law Ising interactions \cite{Cirac1995,Bruzewicz2019,Monroe2019}. Power-law interactions can lead to nontrivial dynamics that need to be examined more closely in the context of quantum many-body scars since they could lead to qualitative changes in the diagram Fig.\ref{Fig_system}a.

Finally, one could alternatively probe the effects of disorder in scar systems with digital quantum simulators, although there are some challenges for currently available hardware \cite{Smith2019}.  However, there has been significant progress in the fidelity of single and two-qubit gates in both ion traps \cite{Wright2019,Nam2020} and SC qubits \cite{Kjaergaard2020}. Furthermore, there has been progress in algorithms for quantum simulation beyond the coherence time of quantum devices \cite{Cirstoiu2020}. As a result, it might be possible to access the physics of scar resonances in some devices in the near future.  Interestingly, in addition to having the potential to model the Rydberg Hamiltonian, digital simulators can directly simulate the strongly interacting limit described by the PXP model using Toffoli gates that effectively impose the kinematic constraint on sets of three contiguous qubits.

The results presented in this work invite further investigation into the non-ergodic nature of scar systems along several directions. For example,  the rigidity we found in the spatio-temporal correlations of the PXP model poses intriguing questions regarding a possible framework for understanding quantum scar systems as realizations of continuous time translation symmetry breaking in static (undriven) systems. It would also be interesting to explore further the transitions at strong disorder that eventually lead to the fully thermal and constrained MBL phases.  Experimentally, our findings could be used to devise strategies to calibrate quantum simulators that suffer from spatial imperfections such as in SC qubits. 

Finally, note that while our work has focused on the one dimensional Rydberg model as a case study, scar resonances can arise in other models and also in higher dimensions. This is clear from the derivation of Eq.\ref{Eq_A0A0_ave}, which did not make assumptions about the specific model at hand. The results in this work thus apply quite generally to other systems and consequently lay the groundwork to study other aspects of non-ergodic dynamics arising from scar states in the presence of disorder.

\acknowledgements

We are grateful to S. Kolkowitz for useful discussions. Research was supported by a QIS Award funded by U.S. Department of Energy, Office of Science, Basic Energy Sciences, at the University of Wisconsin under Award No. DE-SC0019449 and at Argonne National Laboratory. This work was performed in part at the Aspen Center for Physics, which is supported by National Science Foundation grant PHY-1607611. We gratefully acknowledge the computing resources provided on Bebop, a high-performance computing cluster operated by the Laboratory Computing Resource Center at Argonne National Laboratory.



\appendix

\section{Magnetization dynamics in the clean limit}\label{App_0}

In this appendix, we derive the magnetization of the deformed Rydberg model in the strongly interacting limit. Since the $\mathbb{Z}_2$ state is approximately spanned by optimized scar states, we can write the initial state in the form $\ket{\Psi(0)}=\sum_{l\in \Lambda_s}\psi_l \ket{\phi^{(0)}_l},$ where $\Lambda_s$ the set of indices that label optimized scars and $\{\ket{\phi^{(0)}_l}\}$ is the set of eigenstates of the clean system. Since the time-evolved state is given by $\ket{\Psi(t)}=\sum_{l\in\Lambda_s}e^{-iE^{(0)}_l t}\psi_l\ket{\phi^{(0)}_l},$ the magnetization takes the form
\begin{equation}
\widetilde{\mathcal{M}}_a(\pi,t)=\sum_{l,l'\in\Lambda_s}e^{i\omega_{l'l}t}\psi^{*}_{l'}\psi_{l}\bra{\phi^{(0)}_{l'}}\widetilde{\sigma}^a_{\pi}\ket{\phi^{(0)}_l},\label{App_ScarM}
\end{equation}
where we defined $\omega_{l'l}=E^{(0)}_{l'}-E^{(0)}_{l}.$ Thus, the problem reduces to determining the functional form of $\bra{\phi^{(0)}_{l'}}\widetilde{\sigma}^a_{\pi}\ket{\phi^{(0)}_l}.$ As was found in \cite{Choi2019}, angular momentum operators can be defined within the scar subspace as 
\begin{subequations}
\begin{eqnarray}
S_x & = & \frac{1}{2\eta}\sum_{r}\mathcal{P}_s\left(\sigma^x_r+\delta \sigma^x_r\right)\mathcal{P}_s    ,\\
S_y & = & \frac{1}{2\eta}\sum_{r}(-1)^r\mathcal{P}_s\left(\sigma^y_r+\delta \sigma^y_r\right)\mathcal{P}_s, \\
S_z & = & -i[S_x,S_y],
\end{eqnarray}\label{Eq_Sxyz}
\end{subequations}
\noindent where $\mathcal{P}_s=\sum_{l\in\Lambda_s}\ket{\phi^{(0)}_l}\bra{\phi^{(0)}_l}$ is the projector into the subspace of optimized scars, and $\delta \sigma^a_r=\frac{2\delta J_R}{\Omega}\left(\sigma^z_{r+2}+\sigma^z_{r-2}\right)\sigma^a_r.$ These operators approximately satisfy angular momentum commutation relations within the subspace of scar states, so that the $L+1$ states span the tower of maximum total angular momentum $S=L/2.$ Now, note that $S_{x}$ and $S_{y}$ are slight deformations of the operators $\sigma^x_{0}$ and $\sigma^y_{\pi},$ respectively, since $\delta J_R/\Omega$ is small. As a result, we approximate
\begin{eqnarray}
    \bra{\phi^{(0)}_{l\in\Lambda_s}}S_{x}\ket{\phi^{(0)}_{l'\in\Lambda_s}}&\approx&\frac{1}{2\eta}\bra{\phi^{(0)}_{l\in\Lambda_s}}\widetilde{\sigma}^{x}_0\ket{\phi^{(0)}_{l'\in\Lambda_s}},\nonumber\\
    \bra{\phi^{(0)}_{l\in\Lambda_s}}S_{y}\ket{\phi^{(0)}_{l'\in\Lambda_s}}&\approx&\frac{1}{2\eta}\bra{\phi^{(0)}_{l\in\Lambda_s}}\widetilde{\sigma}^{y}_{\pi}\ket{\phi^{(0)}_{l'\in\Lambda_s}}.\nonumber
\end{eqnarray}
We emphasize that, while we have disregarded the deformation term proportional to $\delta J_R/\Omega$ in these angular momentum operators, the eigenstates $\{\ket{\phi^{(0)}_l}\}$ are obtained \textit{with} the deformed Hamiltonian. This reduces the deviation from the ideal $\mathfrak{su}(2)$ algebra found in \cite{Choi2019}.

Next, the optimization of scar states implemented in \cite{Choi2019} is designed so that the $L+1$ states $\{\ket{K_n}\}$ obtained by the Forward Scattering Approximation (FSA) proposed in \cite{Turner2018a} become exact eigenstates of $S_z$ so that $\bra{K_n}S_{z}\ket{K_{n'}}\approx m_n\delta_{nn'},$ where $m_n=-L/2,\ldots,L/2.$  As a result, the projection operator into the scar subspace can be written in the unitarily equivalent form $\mathcal{P}_s=\sum_{n}\ket{K_n}\bra{K_n},$ since the $\{\ket{K_n}\}$ must span this subspace. These FSA states are constructed by repeated application of the operator $H^{+}=\sum_r(\sigma^x_r+\delta \sigma^x_r)+i \sum_r(-1)^r(\sigma^y_r+\delta \sigma^y_{r})$ on the $\mathbb{Z}_2$ state \cite{Choi2019}. Since $[\sigma^z_{\pi},H^+]=2H^{+}$ and $\ket{\mathbb{Z}_2}$ is an eigenstate of $\sigma^z_\pi,$ then $\bra{K_n}\sigma^z_\pi\ket{K_{n'}}= 2 m_n\delta_{nn'}.$ Thus, we can write approximately
\begin{equation}
    S_{z}\approx\frac{1}{2}\mathcal{P}_s\widetilde{\sigma}^{z}_\pi\mathcal{P}_s.
\end{equation}
Having established the approximate $\mathfrak{su}(2)$ algebra satisfied by  $\sigma^{x}_{0}$ and $\sigma^{y,z}_{\pi},$ we now express the latter in the form
\begin{eqnarray}
    \mathcal{P}_s\widetilde{\sigma}^z_{\pi}\mathcal{P}_s&=&\mathcal{P}_s\left(\sigma^{+}_s+\sigma^{-}_s\right)\mathcal{P}_s\\
    \mathcal{P}_s\widetilde{\sigma}^y_{\pi}\mathcal{P}_s&=&\eta\mathcal{P}_s\left(\sigma^{+}_{s}-\sigma^{-}_{s}\right)\mathcal{P}_s.
\end{eqnarray} 
Here, the $\widetilde{\sigma}^{\pm}_{s}$ act as ladder operators on the scar eigenstates:  $\bra{\phi^{(0)}_{l'\in\Lambda_s}}\sigma^{\pm}_s\ket{\phi^{(0)}_{l\in\Lambda_s}}\propto \delta_{l',l_{\pm 1}}$  where the notation $l_{\pm 1}$ refers to the index of the scar state that is spectrally separated by $\pm \omega_{\text{scar}}$ from the $l$-th scar state. Thus, the matrix elements $\bra{\phi^{(0)}_{l'\in \Lambda_s}}\widetilde{\sigma}^a_{\pi}\ket{\phi^{(0)}_{l\in\Lambda_s}}$ precisely pick out the phases that oscillate at the frequency $\omega_{l_{\pm}l}=\omega_{\text{scar}}.$ We then obtain the magnetization components
\begin{subequations}
\begin{eqnarray}
    \widetilde{\mathcal{M}}_y(\pi,t)&\approx&\eta L\sin\left(\omega_{\text{scar}}t\right),\\
    \widetilde{\mathcal{M}}_z(\pi,t)&\approx&-L \cos\left(\omega_{\text{scar}}t\right),
\end{eqnarray}
\end{subequations}
To arrive at this expression, we have used that $\sum_{l\in\Lambda'_s}2\text{Re}\left(\psi^*_{l+1}\psi_l \bra{\phi^{(0)}_{l_{+1}}}\sigma^+_s\ket{\phi^{(0)}_l}\right)\approx\widetilde{\mathcal{M}}_z(\pi,0)=-L,$ where $\Lambda'_s$ excludes the highest-energy scar state. These are the expressions presented in Eq.\ref{Eq_MScarEx} of the main text.

\section{Reordering of basis}\label{App_Reorg}

In this appendix, we discuss how to re-order the energy eigenstates into towers. Suppose we take an energy eigenstate $\ket{\phi^{(0)}_{m_1}}$ from the bottom of the spectrum and we apply the ladder operator once. We use the operators defined in the main text
\begin{equation}
    \sigma^{\pm}_{\text{s}}=\frac{1}{2}\left(\widetilde{\sigma}^z_{\pi}\mp i\eta^{-1} \widetilde{\sigma}^y_{\pi}\right).
\end{equation}
As we discussed in Appendix \ref{App_ScarM}, these are approximate ladder operators within the space of optimized scar states. Here, however, we will examine their behavior in the full PXP Hilbert space. In general, applying $\sigma^{+}_s$ on any given state can result in a linear combination over all eigenstates of the clean system. However, numerically one finds that this linear combination is highly concentrated at a particular state $\ket{\phi^{(0)}_{m_2}}$ of higher energy than $\ket{\phi^{(0)}_{m_1}}.$ If we subsequently compute $\sigma^{+}_s\ket{\phi^{(0)}_{m_2}},$ one again finds that this state is largely peaked at another eigenstate $\ket{\phi^{(0)}_{m_3}}$ with higher energy than $\ket{\phi^{(0)}_{{m_1}}}$ and $\ket{\phi^{(0)}_{{m_2}}}.$ By repeating this procedure, we eventually arrive at a state $\ket{\phi^{(0)}_{m_{max}}}$ for which the further application of the operator does not produce a peaked distribution. Thus, $\sigma^{+}_s$ acts like a ladder operator for the set of states $\{\ket{\phi^{(0)}_{m_i}}\},$ even in cases where the states do not belong to the scar subspace. By iteratively applying this procedure to all eigenstates, we can systematically group all energy eigenstates into sets $\{\ket{\phi^{(0)}_{Jm}}\},$ where $J$ labels the set and $m=1,\ldots,\mathcal{D}_J$ labels the state within that set. In this notation, we can include the scar space in the set $J=1,$ since this set is generated by starting with the ground state. We found that a small number of states do not have strong matrix elements with any other state in the basis, so they form sets by themselves according to our algorithm. Clearly, such states do not contribute to the magnetization. 

This is the set of steps we took to re-order the basis of energy eigenstates. For example, the case $L=8$ presented in the main text leads to $13$ towers of sizes: $\mathcal{D}_1=9,$ $\mathcal{D}_2=\mathcal{D}_3=7,$ $\mathcal{D}_4=\mathcal{D}_5=6,$ $\mathcal{D}_{7-13}=1.$

\section{Disorder-averaged amplitude dynamics}\label{App_AA}

In this section, we will calculate the disorder-averaged quantity $[\mathcal{A}^*_{l'}(t)\mathcal{A}_{l}(t)],$ where the amplitudes $\mathcal{A}_{l}(t)$ determine the time evolution of the system
$\ket{\Psi(t)}=\sum_{l}\mathcal{A}_{l}(t)e^{-i E^{(0)}_{l} t}\ket{\phi^{(0)}_{l}},$
where $\ket{\phi^{(0)}_l}$ is the $l$-th eigenstate of the clean Hamiltonian with energy $E^{(0)}_l.$ 
The Schrodinger equation for these amplitudes is given by
\begin{equation}
    i\frac{d}{dt}\mathcal{A}_{l}(t)=\sum_{l'}\bra{\phi^{(0)}_{l}}\hat{W}\ket{\phi^{(0)}_{l'}}\mathcal{A}_{l'}(t)e^{i \omega_{ll'}t},
\end{equation}
where $\omega_{ll'}=E^{(0)}_l-E^{(0)}_{l'}.$ We will proceed by first calculating the dynamics of the amplitudes corresponding to optimized scars $l\in \Lambda_s$ and afterwards obtain the dynamics of the remaining amplitudes $l\not\in \Lambda_s.$

\subsubsection{Dynamics of the tower of optimized scars}

We begin by calculating the dynamics of the tower of maximum angular momentum, which consists of the optimized scar states:
\begin{equation}
     i \frac{d}{dt}\mathcal{A}_{l\in\Lambda_s}(t)=\sum_{l'}\bra{\phi^{(0)}_{l}}\hat{W}\ket{\phi^{(0)}_{l'}}e^{i\omega_{ll'}t}\mathcal{A}_{l'}(t).
\end{equation}
The right-hand side of this equation in general includes all possible states in the Hilbert space. However, for weak disorder, we expect that only a subset of states in the spectral neighbourhood of the state $\ket{\phi^{(0)}_{l\in\Lambda_s}}$ will be relevant in the dynamics of $\mathcal{A}_{l\in\Lambda_s}(t).$ We thus truncate the sum to obtain
\begin{eqnarray}
    i \frac{d}{dt}\mathcal{A}_{l\in\Lambda_s}(t)&\approx&\bra{\phi^{(0)}_{l}}\hat{W}\ket{\phi^{(0)}_{l}}\mathcal{A}_{l}(t)\nonumber\\
    &&\quad\quad+\sum_{\substack{l'\not\in\Lambda_s\\\vert\omega_{ll'}\vert<\Delta}}\bra{\phi^{(0)}_{l}}\hat{W}\ket{\phi^{(0)}_{l'}}e^{i\omega_{ll'}t}\mathcal{A}_{l'}(t).\nonumber
\end{eqnarray}
Given that the number of states in the spectral vicinity of the $l'$-th optimized scar state grows exponentially in the system size, we expect that $\mathcal{A}_{l\in\Lambda_s}(t)$ will decay into those states irreversibly.  The amplitudes to which the initial state decays are themselves ruled by the equation
\begin{eqnarray}
    &&i \frac{d}{dt}\mathcal{A}_{l\not\in\Lambda_s}(t)=\sum_{\vert\omega_{ll'}\vert<\Delta}\bra{\phi^{(0)}_{l}}\hat{W}\ket{\phi^{(0)}_{l'}}\mathcal{A}_{l'}(t)e^{i\omega_{ll'}t},\nonumber
\end{eqnarray}
which can be written in the integral form
\begin{equation}
    \mathcal{A}_{l\not\in\Lambda_s}(t)=(-i)\int^{t}_0d\tau\sum_{\vert\omega_{ll'}\vert<\Delta}\bra{\phi^{(0)}_{l}}\hat{W}\ket{\phi^{(0)}_{l'}}\mathcal{A}_{l'}(\tau)e^{i\omega_{ll'}\tau}.
\end{equation}
When this is plugged into the equation for $\mathcal{A}_{l\in \Lambda_s}(t),$ we obtain
\begin{widetext}
\begin{eqnarray}
    i \frac{d}{dt}\mathcal{A}_{l\in\Lambda_s}(t)&\approx&\bra{\phi^{(0)}_{l}}\hat{W}\ket{\phi^{(0)}_{l}}\mathcal{A}_{l}(t)-i\sum_{\substack{l'\not\in\Lambda_s\\\vert\omega_{ll'}\vert<\Delta}} \bra{\phi^{(0)}_{l}}\hat{W}\ket{\phi^{(0)}_{l'}}e^{i\omega_{ll'}t}\int^{t}_0d\tau\sum_{\vert \omega_{l'l''}\vert<\Delta}\bra{\phi^{(0)}_{l'}}\hat{W}\ket{\phi^{(0)}_{l''}}\mathcal{A}_{l''}(\tau)e^{i\omega_{l'l''}\tau},\label{appEq_1}\\
    &\approx&\bra{\phi^{(0)}_{l}}\hat{W}\ket{\phi^{(0)}_{l}}\mathcal{A}_{l}(t)-i\int^{t}_0d\tau \left\{\sum_{\substack{l'\not\in\Lambda_s\\\vert\omega_{ll'}\vert<\Delta}} \vert\bra{\phi^{(0)}_{l}}\hat{W}\ket{\phi^{(0)}_{l'}}\vert^2e^{-i\omega_{ll'}(\tau-t)}\right\}\mathcal{A}_{l}(\tau),\label{appEq_2}\\
    &\approx&\bra{\phi^{(0)}_{l}}\hat{W}\ket{\phi^{(0)}_{l}}\mathcal{A}_{l}(t)-i\mathcal{A}_{l}(t)\int^{t}_0d\tau \left\{\sum_{\substack{l'\not\in\Lambda_s\\\vert\omega_{ll'}\vert<\Delta}} \vert\bra{\phi^{(0)}_{l}}\hat{W}\ket{\phi^{(0)}_{l'}}\vert^2e^{-i\omega_{ll'}(\tau-t)}\right\},\label{appEq_3}\\
    &=&z_{l\in \Lambda_s}(t)\mathcal{A}_{l\in \Lambda_s}(t),\label{appEq_4}
\end{eqnarray}
\end{widetext}
where $\,z_{l\in \Lambda_s}(t)=\bra{\phi^{(0)}_{l}}\hat{W}\ket{\phi^{(0)}_{l}}-i\int^{t}_0d\tau \sum_{\substack{l'\not\in\Lambda_s\\\vert\omega_{ll'}\vert<\Delta}} \vert\bra{\phi^{(0)}_{l}}\hat{W}\ket{\phi^{(0)}_{l'}}\vert^2e^{-i\omega_{ll'}(\tau-t)}.$
To obtain Eq.~(\ref{appEq_4}), we made two approximations which are typically used, for example, in the study of the decay of a discrete level into a continuum. The first approximation was made in going from Eq.\ref{appEq_1} to Eq.\ref{appEq_2}, where we dropped matrix elements that do not connect states with the scar state $\ket{\phi^{(0)}_{l\in\Lambda_s}},$ as these are not dominant in the process of this initial state decaying irreversibly. The second approximation was made in going from Eq.\ref{appEq_2} to Eq.\ref{appEq_3}, where we used that the factor in curly brackets is peaked at $t=\tau,$  so we can evaluate it at $\mathcal{A}(\tau=t)$ and pull it out of the integral.  We thus obtain an equation exclusively for $\mathcal{A}_{l\in\Lambda_s}(t)$ which leads to the solution
\begin{equation}
    \mathcal{A}_{l\in\Lambda_s}(t)=\psi_n e^{-i \int^t_0 z_{l}(\tau')d\tau'}.
\end{equation}
Now, in order to perform the disorder averaging, it is convenient to re-write this so that the dependence on the random variables of the potential are made explicit.
We define the vectors $\mathbf{h}^T=\{h_{a(1)}(r(1)),\ldots,h_{a(3L)}(r(3L))\}$ and $\rho^T_{mn}=\{\rho^{1}_{mn},\ldots,\rho^{3L}_{mn}\}$ where $(\rho_{mn})_{l}=\bra{\phi^{(0)}_n}\sigma^{a(l)}_{r(l)}\ket{\phi^{(0)}_n}$, with $\quad l=1, \ldots, 3L$,  $a(l)=(l-1)\text{mod} 3$,  and $r(l)=\left\lfloor \frac{l-1}{3}\right\rfloor$. The matrix elements of the disorder operator can then be written in the form
\begin{eqnarray}
    \bra{\phi^{(0)}_{l}}\hat{W}\ket{\phi^{(0)}_{l'}}&=&\sum_{ra}h_{a}(r)\bra{\phi^{(0)}_l}\sigma^{a}_{r}\ket{\phi^{(0)}_{l'}}\nonumber\\
    &=& \rho^T_{ll'} \cdot \textbf{h},\\
    \vert \bra{\phi^{(0)}_{l}}\hat{W}\ket{\phi^{(0)}_{l'}}\vert^2&=&\left(\textbf{h}^{T} \cdot \rho^{*}_{ll'}\right) \left(\rho^T_{ll'} \cdot \textbf{h}\right)\nonumber\\
    &=&\textbf{h}^T \cdot \left(\rho^*_{ll'}\cdot\rho^{T}_{ll'}\right)\cdot\textbf{h},
\end{eqnarray}
so we finally have the expression
\begin{equation}
\int^{t}_0z_{l\in\Lambda_s}(\tau)d\tau=t\,\rho^T_{ll}\cdot\textbf{h}-i\textbf{h}^T \cdot G_{l}(t,0)\cdot\textbf{h},
\end{equation}
where $G_{l\in\Lambda_s}(t_1,t_2)=\sum_{\substack{l\neq l ',l'\not\in\Lambda_s\\\vert\omega_{ll'}\vert<\Delta}}\left(\rho^*_{ll'}\cdot\rho^{T}_{ll'}\right)f_{ll'}(t_1,t_2)$ and $f_{mn}(t_1,t_2)=i\frac{(t_1-t_2)}{\omega_{mn}}+\frac{e^{i\omega_{mn}t_2}-e^{i\omega_{mn}t_1}}{\omega_{mn}^2}.$ We defined this function in terms of two temporal arguments because it will arise in this form when we calculate the dynamics of the lower towers. The amplitude can then be written explicitly in the form
\begin{equation}
    \mathcal{A}_{l\in\Lambda_s}(t)\approx\psi_l e^{-i t \,\rho_{ll}\cdot\textbf{h}}e^{-\textbf{h}^T \cdot G_{l}(t,0)\cdot\textbf{h}}.
\end{equation}
With this expression at hand, we can calculate $[\mathcal{A}^*_{l'}(t)\mathcal{A}_{l}(t)].$ To do this, we use the Gaussian distribution $ P(\textbf{h})=\frac{1}{\left(\pi W^2/6\right)^{3L/2}}e^{-\frac{6}{W^2}\textbf{h}^2}.$ This distribution has the same first and second moments of the box distribution we used numerically. We thus have
\begin{eqnarray}
    \left[\mathcal{A}^*_{l'}(t)\mathcal{A}_{l}(t)\right]
    &\approx&\frac{\psi^*_{l'}\psi_l}{\left(\pi W^2/6\right)^{3L/2}}\int_{\mathbb{R}^{3L}} d \mathbf{\textbf{h}}e^{-\frac{6}{W^2}\mathbf{\textbf{h}}^2}\\
    &&\qquad \times e^{it\,\left(\rho_{l'l'}-\rho_{ll}\right)\cdot\textbf{h}-\textbf{h}^T \cdot\left( G^*_{l'}(t,0)+G_{l}(t,0)\right)\cdot\textbf{h}}\nonumber\\
    &=&\psi^*_{l'}\psi_l\sqrt{\det\left\{\alpha_{l'l}(t)\right\}}\\
    &&\quad\times e^{-\frac{W^2t^2}{24}\left(\rho_{l'l'}-\rho_{ll}\right)\cdot \alpha_{l'l}(t) \cdot \left(\rho_{l'l'}-\rho_{ll}\right)}\nonumber
\end{eqnarray}
where $\alpha_{l'l}(t)=\left(\mathbb{I}+\frac{W^2}{6}\left(G^*_{l'}(t,0)+G_{l}(t,0)\right)\right)^{-1}.$ This is the expression Eq.~(\ref{Eq_A0A0_ave}) discussed in the main text.

\subsubsection{Dynamics of $J\neq 1$ towers}

We now move on to examine the dynamics of the other towers. The main difference with the tower of optimized scars is the initial condition. Again, the equation is given by
\begin{equation}
    i \frac{d}{dt}\mathcal{A}_{l\not \in \Lambda_s}(t)=\sum_{l'}\bra{\phi^{(0)}_{l}}\hat{W}\ket{\phi^{(0)}_{l'}}A_{l'}(t)e^{i\omega_{ll'}t}.
\end{equation}
Once the amplitudes for the lower towers begin to grow from zero, there will be tunneling from these states to all other spectrally near-by states. The tower of optimized scars thus serves as a source for the lower towers and the spectrally surrounding states serve as a sink into which the amplitudes $\mathcal{A}_{l\not \in \Lambda_s}(t)$ decay irreversibly. Similar to the approach used for the tower of maximal angular momentum, we again express the amplitude of the states to which the amplitude $
\mathcal{A}_{l}(t)$ decays in the integral form $\mathcal{A}_{l'}(t)=(-i)\int^{t}_0d\tau\sum_{\vert\omega_{l'l''}\vert<\Delta}\bra{\phi^{(0)}_{l'}}\hat{W}\ket{\phi^{(0)}_{l''}}\mathcal{A}_{l''}(\tau)e^{i\omega_{l'l''}\tau}.$ We can then write the equation for $\mathcal{A}_{l\not \in \Lambda_s}(t)$ as
\begin{widetext}
\begin{small}
\begin{eqnarray}
    i \frac{d}{dt}\mathcal{A}_{l\not\in\Lambda_s}(t)&=&\bra{\phi^{(0)}_{l}}\hat{W}\ket{\phi^{(0)}_{l}}\mathcal{A}_{l}(t)+\bra{\phi^{(0)}_{l}}\hat{W}\ket{\phi^{(0)}_{n_0(l)}}e^{it\omega_{ln_0(l)}}\mathcal{A}_{n_0(l)}(t)-i\sum_{\substack{l'\not\in\Lambda_s,l'\neq l\\\vert\omega_{ll'}\vert<\Delta}} \bra{\phi^{(0)}_{l}}\hat{W}\ket{\phi^{(0)}_{l'}}e^{i\omega_{ll}t}\mathcal{A}_{l'}(t)\label{appEq_5}\\
    &=&\bra{\phi^{(0)}_{l}}\hat{W}\ket{\phi^{(0)}_{l}}\mathcal{A}_{l}(t)+\bra{\phi^{(0)}_{l}}\hat{W}\ket{\phi^{(0)}_{n_0(l)}}e^{it\omega_{ln_0(l)}}\mathcal{A}_{n_0(l)}(t)\label{appEq_6}\\
    &&\qquad\qquad\qquad\qquad\qquad\qquad-i\sum_{\substack{l'\not\in\Lambda_s,l'\neq l\\\vert\omega_{ll'}\vert<\Delta}} \bra{\phi^{(0)}_{l}}\hat{W}\ket{\phi^{(0)}_{l'}}e^{i\omega_{ll'}t}\int^{t}_0d\tau\sum_{\vert\omega_{l'l''}\vert<\Delta}\bra{\phi^{(0)}_{l'}}\hat{W}\ket{\phi^{(0)}_{l''}}\mathcal{A}_{l''}(\tau)e^{i\omega_{l'l''}\tau}\nonumber\\
    &=&\bra{\phi^{(0)}_{l}}\hat{W}\ket{\phi^{(0)}_{l}}\mathcal{A}_{l}(t)+\bra{\phi^{(0)}_{l}}\hat{W}\ket{\phi^{(0)}_{n_0(l)}}e^{it\omega_{ln_0(l)}}\mathcal{A}_{n_0(l)}(t)\label{appEq_7}\\
    &&\qquad\qquad\qquad\qquad\qquad\qquad-i\int^{t}_0d\tau\left\{\sum_{\substack{l'\not\in\Lambda_s,l'\neq l\\\vert\omega_{ll'}\vert<\Delta}}\sum_{\vert\omega_{l'l''}\vert<\Delta} \bra{\phi^{(0)}_{l}}\hat{W}\ket{\phi^{(0)}_{l'}}\bra{\phi^{(0)}_{l'}}\hat{W}\ket{\phi^{(0)}_{l''}}e^{i\omega_{ll'}t-i\omega_{l''l'}\tau}\right\}\mathcal{A}_{l''}(\tau)\nonumber
\end{eqnarray}
\end{small}
\end{widetext}
Now, the dominant terms inside the curly brackets will occur for $J=J''$ and $t\sim \tau.$ By keeping these matrix elements and evaluating the amplitude at $t=\tau,$ we obtain the equation
\begin{equation}
    i \frac{d}{dt}\mathcal{A}_{l\not\in \Lambda_s}(t)\approx\bra{\phi^{(0)}_{l}}\hat{W}\ket{\phi^{(0)}_{n_0(l)}}e^{it\omega_{ln_0(l)}}\mathcal{A}_{n_0(l)}(t)+z_{l}(t)\mathcal{A}_{l}(t)
\end{equation}
where in this case $z_{l\not\in \Lambda_s}(t)=\bra{\phi^{(0)}_{l}}\hat{W}\ket{\phi^{(0)}_{l}}-i\int^{t}_0d\tau\sum_{\substack{l'\not\in\Lambda_s,l'\neq l\\\vert\omega_{ll'}\vert<\Delta}} e^{i(t-\tau)\omega_{ll'}}\vert\bra{\phi^{(0)}_{l}}\hat{W}\ket{\phi^{(0)}_{l'}}\vert^2.$ The solution for the amplitude is 
\begin{equation}
    \mathcal{A}_{l \not \in \Lambda_s}(t)\approx-i \bra{\phi^{(0)}_{l}}\hat{W}\ket{\phi^{(0)}_{n_0(l)}}\int^{t}_0 e^{-i\mathcal{F}_l(t,\tau)}\mathcal{A}_{n_0(l)}(\tau)d\tau,
\end{equation}
where we defined
$\mathcal{F}_l(t,\tau)=-\tau\omega_{l,n_0(l)}+(t-\tau)\,\rho_{ll}\cdot\textbf{h}-i\textbf{h}^T \cdot G_{l}(t,\tau)\cdot\textbf{h}.$ This is the expression for the amplitude we used in the main text. With this expression at hand, the disorder averaged quantity $\left[\mathcal{A}^*_{l'}(t)\mathcal{A}_{l}(t)\right]$ we require is
\begin{eqnarray}
    &&\left[\mathcal{A}^*_{l'}(t)\mathcal{A}_{l}(t)\right]\approx\sum_{ll'}\psi^*_{n_0(l')}\psi_{n_0(l)}\left(\rho^{*}_{l'n_0(l')}\right)_{n'}\left(\rho_{l,n_0(l)}\right)_n\nonumber\\
    &&\qquad\qquad\qquad \times \int d^2\vec{\tau} \,e^{-i\tau_2\omega_{l'n_0(l')}+i\tau_1\omega_{ln_0(l)}}\nonumber\\
    &&\qquad\qquad\qquad\qquad \times\left[\textbf{h}_{n'}\textbf{h}_{n}e^{-\frac{1}{2}\textbf{h}^T\cdot \Gamma_{l'l}(t,\vec{\tau})\cdot\textbf{h}  }e^{i R_{l'l}(t,\vec{\tau})\cdot \textbf{h}}\right],\nonumber
\end{eqnarray}
where $\Gamma_{l'l}(t,\vec{\tau})=2(G^{\dagger}_{l'}(t,\tau_1)+G_{l}(t,\tau_2)+G^{\dagger}_{n_0(l')}(\tau_1,0)+G_{n_0(l)}(\tau_2,0),$ and $R_{l'l}(t,\vec{\tau})=\rho_{l'l'}(t-\tau_2)-\rho_{ll}(t-\tau_1)+\rho_{n_0(l')n_0(l')}\tau_2-\rho_{n_0(l),n(l)}\tau_1.$ Note that we can write
\begin{eqnarray}
    &&\left[\textbf{h}_{n'}\textbf{h}_{n}e^{-\frac{1}{2}\textbf{h}^T\cdot \Gamma_{l'l}(t,\vec{\tau})\cdot\textbf{h}  }e^{i R_{l'l}(t,\vec{\tau})\cdot \textbf{h}}\right]\\
    &&=-\frac{\partial}{\partial (R_{l'l})_{n}}\frac{\partial}{\partial (R_{l'l})_{n'}}\left[e^{-\frac{1}{2}\textbf{h}^T\cdot \Gamma_{l'l}(t,\vec{\tau})\cdot\textbf{h}  }e^{i R_{l'l}(t,\vec{\tau})\cdot \textbf{h}}\right],\nonumber
\end{eqnarray}
where the derivative is only with respect to diagonal matrix elements of the Pauli operators, so it does not act on the $\Gamma_{l'l}(t,\vec{\tau})$ matrix. Thus, we only need disorder average the quantity
\begin{eqnarray}
    &&\left[e^{-\frac{1}{2}\textbf{h}^T\cdot \Gamma_{l'l}(t,\vec{\tau})\cdot\textbf{h}  }e^{i R_{l'l}(t,\vec{\tau})\cdot \textbf{h}}\right]=\nonumber\\
    &&\frac{1}{\left(\pi W^2/6\right)^{3L/2}}\int_{\mathbb{R}^{3L}} d \textbf{h} e^{-\frac{6}{W^2}\textbf{h}^2} e^{-\frac{1}{2}\textbf{h}^T\cdot \Gamma_{l'l}(t,\vec{\tau})\cdot\textbf{h}  }e^{i R_{l'l}(t,\vec{\tau})\cdot \textbf{h}},\nonumber\\
    &&=\sqrt{\det\left\{\beta_{l'l}(t,\vec{\tau})\right\}}e^{-\frac{W^2}{24}R^T_{l'l}(t,\tau,\tau')\cdot \beta_{l'l}(t,\vec{\tau}) \cdot R_{l'l}(t,\vec{\tau}) },\nonumber\\
    &&=\mathcal{J}_{l'l}(t,\vec{\tau}).
\end{eqnarray}
In these expressions, we defined $\beta_{l'l}(t,\vec{\tau})=\left(\mathbb{I}+\frac{W^2}{12} \Gamma_{l'l}(t,\vec{\tau})\right)^{-1}$. The first and second derivative of this expression leads to
\begin{eqnarray}
    &&\frac{\partial}{\partial (R_{l'l})_{n}}\mathcal{J}_{l'l}(t,\vec{\tau})=\mathcal{J}_{l'l}(t,\vec{\tau})\left(-\frac{W^2}{24}\right)\Delta^{n}_{l'l}(t,\vec{\tau})\\
&&\frac{\partial}{\partial (R_{l'l})_{n'}}\frac{\partial}{\partial (R_{l'l})_{n}}\mathcal{J}_{l'l}(t,\vec{\tau})=\\
&&\quad\quad\mathcal{J}_{l'l}(t,\vec{\tau})\left(-\frac{W^2}{24}\right)\left(\left\{\beta_{l'l}(t,\vec{\tau})\right\}^{l}_{l'}+\left\{\beta_{l'l}(t,\vec{\tau})\right\}^{l'}_{l}\right)\nonumber\\
&&\quad\quad\quad\quad\quad\quad+\mathcal{J}_{l'l}(t,\vec{\tau})\left(-\frac{W^2}{24}\right)^2\Delta^{n'}_{l'l}(t,\vec{\tau})\Delta^{n}_{l'l}(t,\vec{\tau}),\nonumber
\end{eqnarray}
with $\Delta^{n}_{l'l}(t,\vec{\tau})=\left\{\beta_{l'l}(t,\vec{\tau})\cdot R_{l'l}(t,\vec{\tau})\right\}_{n}+\left\{R^T_{l'l}(t,\vec{\tau})\cdot\beta_{l'l}(t,\vec{\tau})\right\}_{n}.$ Putting it all together, the final expression is then 
\begin{widetext}
\begin{eqnarray}
&&\left[\mathcal{A}^*_{l'}(t)\mathcal{A}_{l}(t)\right]\approx\psi^*_{n_0(l')}\psi_{n_0(l)}\frac{W^2}{24}\int d^2\vec{\tau} e^{-i\tau_1\omega_{l'n_0(l')}+i\tau_2\omega_{ln_0(l)}}\sqrt{\det\left\{\beta_{l'l}(t,\vec{\tau})\right\}}e^{-\frac{W^2}{24}R^T_{l'l}(t,\vec{\tau})\cdot \beta_{l'l}(t,\vec{\tau}) \cdot R_{l'l}(t,\vec{\tau}) }\\
&& \qquad \quad \times\left(\rho^{\dagger}_{l'n_0(l')}\cdot\beta_{l'l}(t,\vec{\tau})\cdot\rho_{ln_0(l)}+\rho^{T}_{ln_0(l)}\cdot\beta_{l'l}(t,\vec{\tau})\cdot\rho^*_{l'n_0(l')}-\frac{W^2}{24}\left\{\rho^{\dagger}_{l'n_0(l')}\cdot\Delta_{l'l}(t,\vec{\tau})\right\}\left\{\Delta^{T}_{l'l}(t,\vec{\tau})\cdot\rho_{ln_0(l)}\right\}\right)\nonumber
\end{eqnarray}
\end{widetext}
which is the result discussed in the main text.

\section{Temporal correlator}\label{App_BB}

In this appendix, we obtain the expression for temporal correlators discussed in the main text. We can capture the temporal correlations encoded in Eq.~(\ref{Eq_correlator}) by evaluating the spatial Fourier transform  $\widetilde{\mathcal{C}}_a(t)=\sum_{R,r_0}\int dt e^{i\pi R}\mathcal{C}_a(r_0,r_0+R,t)=\bra{\Psi}\left(\mathcal{P}\sigma^a_{\pi}(t)\mathcal{P}\right)\left(\mathcal{P}\sigma^a_{\pi}(0)\mathcal{P}\right)\ket{\Psi}-[\widetilde{\mathcal{M}}_a(\pi,t)]\widetilde{\mathcal{M}}_a(\pi,0).$ By expanding $\ket{\Psi}=\sum_{m}\psi_{J=1,m}\ket{\phi^{(0)}_{J=1,m}},$ where we used the notation $(Jm)$ that identifies towers in the spectrum, the action of $\mathcal{P}\sigma^a_{\pi}(0)\mathcal{P}$ on this state leads to 
\begin{eqnarray}
    \mathcal{P}\widetilde{\sigma}^a_{\pi}(0)\mathcal{P}\ket{\mathbb{Z}_2}&\approx&\sum_{m}\overline{\psi}^a_{J=1,m}\ket{\phi^{(0)}_{J=1,m}}\equiv \ket{\Phi_a(0)},
\end{eqnarray}
where $\overline{\psi}^a_{J=1,m}=\sum_{s,d}\psi_{1,m-sd}\Gamma^{a,sd}_{1,m-s d}$ with $s=\pm 1,$ $d=1,3.$ For convenience, we define $\psi_{1,m},\Gamma^{a,sd}_{1,m}=0$ whenever $m\not \in [1,\mathcal{D}_J].$ Note that we have included matrix elements with $d=3,$ which go beyond the tri-diagonal in the matrix representation of $\sigma^a_{\pi}.$  In contrast to the magnetization, for the calculation of temporal correlators we need matrix elements beyond the tri-diagonal that are small but nevertheless contribute to the final temporal correlator. 

Now, we have that $\bra{\Psi}\left(\mathcal{P}\sigma^a_{\pi}(t)\mathcal{P}\right)\left(\mathcal{P}\sigma^a_{\pi}(0)\mathcal{P}\right)\ket{\Psi}=\bra{\Psi}\left(\mathcal{P}\sigma^a_{\pi}(t)\mathcal{P}\right)\ket{\Phi_a(0)}=\bra{\Psi(t)}\sigma^a_{\pi}(0)\ket{\Phi_a(t)}.$ The state $\ket{\Phi_a(0)}$ is then an un-normalized initial condition on which the unitary evolution operator acts. We can expand this time-evolved state in the form
\begin{eqnarray}
    \ket{\Phi_a(t)}&=&\sum_{Jm}\mathcal{B}^a_{J m}(t)e^{-i E^{(0)}_{Jm} t}\ket{\phi^{(0)}_{Jm}}.
\end{eqnarray}
This is formally the same starting point as for the calculation of the magnetization, implying that $\mathcal{B}^a_{Jm}(t)=\left(\frac{\overline{\psi}^a_{n_0(Jm)}}{\psi_{n_0(Jm)}}\right)\mathcal{A}_{Jm}(t).$ The correlator can thus be written in the form $[\widetilde{\mathcal{C}}_a(t)]=[\bra{\Psi(t)} \widetilde{\sigma}^a_{\pi} \ket{\Phi_a(t)}]-[\widetilde{\mathcal{M}}_a(\pi,t)]\widetilde{\mathcal{M}}_a(\pi,0),$ where
\begin{eqnarray}
[\bra{\Psi(t)} \widetilde{\sigma}^a_{\pi} \ket{\Phi_a(t)}]&\approx&\sum_{Jm}\sum_{sd}\left(\frac{\overline{\psi}^a_{n_0(Jm-sd)}\Gamma^{a,sd}_{Jm-sd}}{\psi_{n_0(Jm)}}\right)\\
&&\quad \times[\mathcal{A}^*_{Jm+s d}(t)\mathcal{A}_{Jm}(t)]e^{it\omega_{Jm+sd,Jm}}. \nonumber
\end{eqnarray}
This is the expression presented in Eq.~(\ref{Eq_TCorr}) of the main text.


%

\end{document}